\documentclass[preprint,3p,sort&compress]{elsarticle}
\pdfoutput=1 
\journal{Nuclear Physics B}

\usepackage{amsmath}
\usepackage{amsfonts}
\usepackage{amssymb}

\usepackage[dvipsnames]{xcolor}
\usepackage[colorlinks]{hyperref}

\usepackage{tikz}
\usetikzlibrary{intersections}
\usetikzlibrary{arrows,decorations.pathmorphing,backgrounds,positioning,fit,petri}
\usetikzlibrary{decorations.markings}
\usetikzlibrary{calc}

\usepackage{graphicx}

\newcommand{\ctx}{\textcolor{Red}{x}}

\newcommand\f[2]{\genfrac{}{}{0pt}{1}{#2}{#1}}

\newcommand{\ep}{\ensuremath{\varepsilon}}%

\begin{document}
\begin{frontmatter}

  \title{
    \vskip-1.5cm{\baselineskip14pt\rm
      \centerline{\normalsize DESY~18-173\hfill ISSN~0418-9833}
      \centerline{\normalsize October 2018\hfill}
    }
    \vskip1.5cm
    Three-loop effective potential of general scalar theory via
    differential equations
  }

  \author[1]{Bernd A. Kniehl}
  \author[1,2]{Andrey F. Pikelner
}
  \author[1]{Oleg L. Veretin}
  
  \address[1]{II.~Institut f\"ur Theoretische Physik, Universit\"at Hamburg,\\
    Luruper Chaussee 149, 22761 Hamburg, Germany}

  \address[2]{
  Bogoliubov Laboratory of Theoretical Physics,
  Joint Institute for Nuclear Research,\\
  141980 Dubna, Russia}

  \begin{abstract}
    We consider the scalar sector of a general renormalizable theory and
    evaluate the effective potential through three loops analytically.
    We encounter three-loop vacuum bubble diagrams with up to two masses and
    six lines, which we solve using differential equations transformed into
    the favorable $\epsilon$ form of dimensional regularization.
    The master integrals of the canonical basis thus obtained are expressed
    in terms of cyclotomic polylogarithms up to weight four.
    We also introduce an algorithm for the numerical evaluation of cyclotomic
    polylogarithms with multiple-precision arithmetic, which is implemented in
    the {\it Mathematica} package {\tt cyclogpl.m} supplied here.
  \end{abstract}

\end{frontmatter}


\section{Introduction}
\label{sec:intro}

The effective potential \cite{Coleman:1973jx,Jackiw:1974cv} plays a very
important role in investigating spontaneous symmetry breaking.
In the Standard Model, it has been a topic of numerous studies over many years
\cite{Sher:1988mj,Lindner:1988ww,Arnold:1991cv,Ford:1992pn,Ford:1992mv},
with partial results at the three- \cite{Martin:2013gka} and four-loop orders
\cite{Martin:2015eia}.
The analysis of the effective potential at the two-loop order leads to the
conclusion that the electroweak vacuum may be stable, critical, or slightly
meatastable up to very high energies of the order of the Planck scale
\cite{Bezrukov:2012sa,Buttazzo:2013uya,Bednyakov:2015sca}.
Such an analysis at the three-loop order would require, apart from four-loop
renormalization group functions
\cite{Bednyakov:2012rb,Bednyakov:2012en,Bednyakov:2013eba,Bednyakov:2015ooa,Zoller:2015tha,Chetyrkin:2016ruf},
also the matching conditions, which are presently known through the two-loop
order only \cite{Kniehl:2015nwa}.
An important step has recently been taken in Ref.~\cite{Martin:2017lqn}, where
the three-loop potential has been studied in a general renormalizable theory
evaluating the loop integrals numerically
\cite{Freitas:2016zmy,Martin:2016bgz,Bauberger:2017nct}.

In this work, we consider the purely scalar sector of a general renormalizable
theory and evaluate the effective potential through three loops analytically.
The integrals that appear in our calculation can have up to two different mass
scales.
In the case of $\mathcal{O}(n)$ symmetry, we reproduce the known result for the
scalar $\varphi^4$ theory with spontaneous symmetry breaking.
This theory is a matter of interest for the study of phase transitions, and its
effective potential has been calculated in a series of papers
\cite{Ford:1991hw,Chung:1997jy}.
The results presented here also reproduce the contribution of the scalar sector
to the effective potential in the Standard Model.





This paper is organized as follows.
In Section~\ref{sec:effect-potent-scal}, we introduce the Lagrangian of the
scalar sector and parametrize the three-loop effective potential in terms of
three-loop master integrals.
In Section~\ref{sec:details}, we discuss the evaluation of the master integrals
with two different mass scales with the help of differential equations.
In Section~\ref{sec:results}, we present our results.
The numerical evaluation of the cyclotomic polylogarithms that appear in our
results is discussed in the appendix.

\section{Effective potential in the scalar theory}
\label{sec:effect-potent-scal}

Let us consider the scalar theory described by the Lagrangian
\begin{equation}
  \label{eq:lag}
  \mathcal{L}_S = \frac{1}{2}(\partial H)^2 + \frac{1}{2}(\partial G)^2
  - \frac{\mu_0^2}{2} H^2 - \frac{\mu_i^2}{2} G_i^2 - \frac{\tau_0}{6} H^3  - \frac{\tau_i}{6} H G_i^2 -
  \frac{\lambda_0}{24} H^4 -\frac{\lambda_i}{12} H^2 G_i^2 - \frac{\lambda_{ij}}{24} G_i^2 G_j^2 \,,
\end{equation}
where $\tau_i$, $\lambda_i$, and $\lambda_{ij}$ are couplings.
In the Standard Model, the scalar part of the potential reads
\begin{equation}
  \mathcal{L} = - m^2\Phi^\dagger\Phi - \lambda(\Phi^\dagger\Phi)^2 \,.
\end{equation}
In the broken phase, it can be parametrized as 
\begin{equation}
  \Phi = \frac{1}{\sqrt{2}} \left( {\phi + H + iG_0 \atop G^+ }\right)  \,,
\end{equation}
where $\phi$ is the vacuum expectation value, $H$ is the Higgs field, and
$G_0$ and $G^\pm$ are the scalar would-be Goldstone bosons.
This parametrization corresponds to the following choice of parameters in
Eq.~(\ref{eq:lag}):
\begin{align}
\label{eq:mmphi}
  \tau_0 = \tau_i = 6 \lambda\phi, \qquad \lambda_0 = \lambda_i = \lambda_{ij} = 6 \lambda,
  \qquad
  \mu_0^2 = m^2 + 3 \lambda\phi^2, \qquad
  \mu_i^2 = m^2 + \lambda\phi^2 \,.
\end{align}

The three-loop contribution to the effective potential of the theory in
Eq.~\eqref{eq:lag} can be schematically represented as the following sum of
Feynman diagrams:

\tikzset{
  H/.style={line width=1pt,draw=Black},
  G/.style={line width=1pt,draw=NavyBlue,dashed}
}


\begin{align}
  V_3 & =
  %
  %
        \frac{\lambda_{0}^2}{16}
        \begin{tikzpicture}[baseline={([yshift=-.5ex]current bounding box.center)},scale=1]
          \useasboundingbox (-1,-1) rectangle (1,1);
          \draw[H] (-0.5,0) circle (0.25);
          \draw[H] (0,0) circle (0.25);
          \draw[H] (0.5,0) circle (0.25);
        \end{tikzpicture}
        + \frac{2}{3}\frac{\lambda_0 \lambda_{i}}{16}
        \begin{tikzpicture}[baseline={([yshift=-.5ex]current bounding box.center)},scale=1]
          \useasboundingbox (-1,-1) rectangle (1,1);
          \draw[H] (-0.5,0) circle (0.25);
          \draw[H] (0,0) circle (0.25);
          \draw[G] (0.5,0) circle (0.25);
          \draw (0.5,0) node {\textcolor{red}{$i$}};
        \end{tikzpicture}
        + \frac{1}{9}\frac{\lambda_{i}^2}{16}
        \begin{tikzpicture}[baseline={([yshift=-.5ex]current bounding box.center)},scale=1]
          \useasboundingbox (-1,-1) rectangle (1,1);
          \draw[H] (-0.5,0) circle (0.25);
          \draw[G] (0,0) circle (0.25);
          \draw[H] (0.5,0) circle (0.25);
          \draw (0,0) node {\textcolor{red}{$i$}};
        \end{tikzpicture}
        \nonumber
  \\
      & + \textcolor{red}{\frac{1}{9}} \frac{\lambda_{i} \lambda_{j}}{16}
        \begin{tikzpicture}[baseline={([yshift=-.5ex]current bounding box.center)},scale=1]
          \useasboundingbox (-1,-1) rectangle (1,1);
          \draw[G] (-0.5,0) circle (0.25);
          \draw[H] (0,0) circle (0.25);
          \draw[G] (0.5,0) circle (0.25);
          \draw (-0.5,0) node {\textcolor{red}{$i$}};
          \draw (0.5,0) node {\textcolor{red}{$j$}};
        \end{tikzpicture}
        + \textcolor{red}{\frac{2(1+2\delta_{ij})}{9}} \frac{\lambda_{i} \lambda_{ij}}{16}
        \begin{tikzpicture}[baseline={([yshift=-.5ex]current bounding box.center)},scale=1]
          \useasboundingbox (-1,-1) rectangle (1,1);
          \draw[H] (-0.5,0) circle (0.25);
          \draw[G] (0,0) circle (0.25);
          \draw[G] (0.5,0) circle (0.25);
          \draw (0,0) node {\textcolor{red}{$i$}};
          \draw (0.5,0) node {\textcolor{red}{$j$}};         
        \end{tikzpicture}
        + \textcolor{red}{\frac{(1+2\delta_{ij})(1+2\delta_{jk})}{9}} \frac{\lambda_{ij} \lambda_{jk}}{16}
        \begin{tikzpicture}[baseline={([yshift=-.5ex]current bounding box.center)},scale=1]
          \useasboundingbox (-1,-1) rectangle (1,1);
          \draw[G] (-0.5,0) circle (0.25);
          \draw[G] (0,0) circle (0.25);
          \draw[G] (0.5,0) circle (0.25);
          \draw (-0.5,0) node {\textcolor{red}{$i$}};
          \draw (0,0) node {\textcolor{red}{$j$}};
          \draw (0.5,0) node {\textcolor{red}{$k$}};
        \end{tikzpicture}
        \label{eq:veff3i1}
  \\
  %
  %
      & + \frac{\lambda_0 \tau_{0}^2}{8}
        \begin{tikzpicture}[baseline={([yshift=-.5ex]current bounding box.center)},scale=1]
          \useasboundingbox (-1,-1) rectangle (1,1);
          \draw[H] (0.5,0.5) -- (0.5,-0.5);
          \draw[H] (-0.25,0) circle (0.25);
          \draw[H] (0.5,-0.5) arc (-90:90:0.5);
          \draw[H] (0.5,0.5) arc (90:270:0.5);
        \end{tikzpicture}
         + \frac{1}{3}\frac{\lambda_i \tau_{0}^2}{8}
        \begin{tikzpicture}[baseline={([yshift=-.5ex]current bounding box.center)},scale=1]
          \useasboundingbox (-1,-1) rectangle (1,1);
          \draw[H] (0.5,0.5) -- (0.5,-0.5);
          \draw[G] (-0.25,0) circle (0.25);
          \draw[H] (0.5,-0.5) arc (-90:90:0.5);
          \draw[H] (0.5,0.5) arc (90:270:0.5); 
          \draw (-0.25,0) node {\textcolor{red}{$i$}};
        \end{tikzpicture}
         + \frac{2}{27}\frac{\lambda_i \tau_i^2}{8}
        \begin{tikzpicture}[baseline={([yshift=-.5ex]current bounding box.center)},scale=1]
          \useasboundingbox (-1,-1) rectangle (1,1);
          \draw[G] (0.5,0.5) -- (0.5,-0.5);
          \draw[H] (-0.25,0) circle (0.25);
          \draw[H] (0.5,-0.5) arc (-90:90:0.5);
          \draw[G] (0.5,0.5) arc (90:270:0.5); 
          \draw (0.25,0) node {\textcolor{red}{$i$}};
        \end{tikzpicture}
\nonumber  \\
      & + \textcolor{red}{\frac{2(1+2\delta_{ij})}{27}} \frac{\lambda_{ij} \tau_j^2}{8}
        \begin{tikzpicture}[baseline={([yshift=-.5ex]current bounding box.center)},scale=1]
          \useasboundingbox (-1,-1) rectangle (1,1);
          \draw[G] (0.5,0.5) -- (0.5,-0.5);
          \draw[G] (-0.25,0) circle (0.25);
          \draw[H] (0.5,-0.5) arc (-90:90:0.5);
          \draw[G] (0.5,0.5) arc (90:270:0.5); 
          \draw (-0.25,0) node {\textcolor{red}{$i$}};
          \draw (0.25,0) node {\textcolor{red}{$j$}};         
        \end{tikzpicture}
        + \frac{1}{9}\frac{\lambda_0 \tau_i^2}{8}
        \begin{tikzpicture}[baseline={([yshift=-.5ex]current bounding box.center)},scale=1]
          \useasboundingbox (-1,-1) rectangle (1,1);
          \draw[G] (0.5,0.5) -- (0.5,-0.5);
          \draw[H] (-0.25,0) circle (0.25);
          \draw[G] (0.5,-0.5) arc (-90:90:0.5);
          \draw[H] (0.5,0.5) arc (90:270:0.5); 
          \draw (0.75,0) node {\textcolor{red}{$i$}};
        \end{tikzpicture}
        + \textcolor{red}{\frac{1}{27}}\frac{\lambda_{i} \tau_j^2}{8}
        \begin{tikzpicture}[baseline={([yshift=-.5ex]current bounding box.center)},scale=1]
          \useasboundingbox (-1,-1) rectangle (1,1);
          \draw[G] (0.5,0.5) -- (0.5,-0.5);
          \draw[G] (-0.25,0) circle (0.25);
          \draw[G] (0.5,-0.5) arc (-90:90:0.5);
          \draw[H] (0.5,0.5) arc (90:270:0.5);
          \draw (-0.25,0) node {\textcolor{red}{$i$}};
          \draw (0.75,0) node {\textcolor{red}{$j$}};         
        \end{tikzpicture}
        \label{eq:veff3i2}
  \\
  %
  %
      & + \frac{\lambda_{0}^2}{48}
        \begin{tikzpicture}[baseline={([yshift=-.5ex]current bounding box.center)},scale=1]
          \useasboundingbox (-1,-1) rectangle (1,1);
          \coordinate (v1) at (0.77,0);
          \coordinate (v2) at (-0.77,0); 
          \draw[H] (v1) arc (-40:-140:1); 
          \draw[H] (v2) arc (140:40:1);
          \draw[H] (0,0) circle (0.77);
        \end{tikzpicture}
        + \frac{2}{3}\frac{\lambda_i^2}{48}
        \begin{tikzpicture}[baseline={([yshift=-.5ex]current bounding box.center)},scale=1]
          \useasboundingbox (-1,-1) rectangle (1,1);
          \coordinate (v1) at (0.77,0);
          \coordinate (v2) at (-0.77,0); 
          \draw[G] (v1) arc (-40:-140:1); 
          \draw[G] (v2) arc (140:40:1);
          \draw[H] (0,0) circle (0.77);
          \draw (0,0) node {\textcolor{red}{$i$}};          
        \end{tikzpicture}
        + \textcolor{red}{\frac{1+2\delta_{ij}}{3}}\frac{\lambda_{ij}^2}{48}
        \begin{tikzpicture}[baseline={([yshift=-.5ex]current bounding box.center)},scale=1]
          \useasboundingbox (-1,-1) rectangle (1,1);
          \coordinate (v1) at (0.77,0);
          \coordinate (v2) at (-0.77,0); 
          \draw[G] (v1) arc (-40:-140:1); 
          \draw[G] (v2) arc (140:40:1);
          \draw[G] (0,0) circle (0.77);
          \draw (0,0.55) node {\textcolor{red}{$i$}};
          \draw (0,-0.55) node {\textcolor{red}{$j$}};         
        \end{tikzpicture}
        \label{eq:veff3i3}
  \\
  %
  %
      & + \frac{\lambda_0 \tau_{0}^2}{8}
        \begin{tikzpicture}[baseline={([yshift=-.5ex]current bounding box.center)},scale=1]
          \useasboundingbox (-1,-1) rectangle (1,1);
          \coordinate (v1) at (90:0.77);
          \coordinate (v2) at (220:0.77);
          \coordinate (v3) at (320:0.77);
          \draw[H] (v1) -- (v2);
          \draw[H] (v1) -- (v3);
          \draw[H] (v3) arc (-40:90:0.77);
          \draw[H] (v1) arc (90:220:0.77);
          \draw[H] (v2) arc (220:320:0.77);
        \end{tikzpicture}
         + \frac{2}{9}\frac{\tau_0 \tau_i \lambda_i}{8}
        \begin{tikzpicture}[baseline={([yshift=-.5ex]current bounding box.center)},scale=1]
          \useasboundingbox (-1,-1) rectangle (1,1);
          \coordinate (v1) at (90:0.77);
          \coordinate (v2) at (220:0.77);
          \coordinate (v3) at (320:0.77);
          \draw[H] (v1) -- (v2);
          \draw[G] (v1) -- (v3);
          \draw[G] (v3) arc (-40:90:0.77);
          \draw[H] (v1) arc (90:220:0.77);
          \draw[H] (v2) arc (220:320:0.77);
          \draw (0.55,0.1) node {\textcolor{red}{$i$}};         
        \end{tikzpicture}
         + \frac{4}{27}\frac{\lambda_i \tau_i^2}{8}
        \begin{tikzpicture}[baseline={([yshift=-.5ex]current bounding box.center)},scale=1]
          \useasboundingbox (-1,-1) rectangle (1,1);
          \coordinate (v1) at (90:0.77);
          \coordinate (v2) at (220:0.77);
          \coordinate (v3) at (320:0.77);
          \draw[G] (v1) -- (v2);
          \draw[G] (v1) -- (v3);
          \draw[H] (v3) arc (-40:90:0.77);
          \draw[H] (v1) arc (90:220:0.77);
          \draw[G] (v2) arc (220:320:0.77);
          \draw (0,0) node {\textcolor{red}{$i$}};         
        \end{tikzpicture}
         + \textcolor{red}{\frac{1+2\delta_{ij}}{27}}\frac{\lambda_{ij} \tau_i \tau_j}{8}
        \begin{tikzpicture}[baseline={([yshift=-.5ex]current bounding box.center)},scale=1]
          \useasboundingbox (-1,-1) rectangle (1,1);
          \coordinate (v1) at (90:0.77);
          \coordinate (v2) at (220:0.77);
          \coordinate (v3) at (320:0.77);
          \draw[G] (v1) -- (v2);
          \draw[G] (v1) -- (v3);
          \draw[G] (v3) arc (-40:90:0.77);
          \draw[G] (v1) arc (90:220:0.77);
          \draw[H] (v2) arc (220:320:0.77);
          \draw (-0.55,0.1) node {\textcolor{red}{$i$}};
          \draw (0.55,0.1) node {\textcolor{red}{$j$}};         
        \end{tikzpicture}
        \label{eq:veff3i4}
  \\
  %
  %
    & + \frac{\tau_{0}^4}{16}
        \begin{tikzpicture}[baseline={([yshift=-.5ex]current bounding box.center)},scale=1]
          \useasboundingbox (-1,-1) rectangle (1,1);
          \coordinate (v1) at (30:0.77);
          \coordinate (v2) at (150:0.77);
          \coordinate (v3) at (210:0.77);
          \coordinate (v4) at (330:0.77);
          \draw[H] (v1) -- (v2);
          \draw[H] (v3) -- (v4);
          \draw[H] (v4) arc (-30:30:0.77);
          \draw[H] (v1) arc (30:150:0.77);
          \draw[H] (v2) arc (150:210:0.77);
          \draw[H] (v3) arc (210:330:0.77);
        \end{tikzpicture}
      + \frac{2}{9}\frac{\tau_{0}^2 \tau_i^2}{16}
        \begin{tikzpicture}[baseline={([yshift=-.5ex]current bounding box.center)},scale=1]
          \useasboundingbox (-1,-1) rectangle (1,1);
          \coordinate (v1) at (30:0.77);
          \coordinate (v2) at (150:0.77);
          \coordinate (v3) at (210:0.77);
          \coordinate (v4) at (330:0.77);
          \draw[G] (v1) -- (v2);
          \draw[H] (v3) -- (v4);
          \draw[H] (v4) arc (-30:30:0.77);
          \draw[G] (v1) arc (30:150:0.77);
          \draw[H] (v2) arc (150:210:0.77);
          \draw[H] (v3) arc (210:330:0.77);
          \draw (0,0.55) node {\textcolor{red}{$i$}};
        \end{tikzpicture}
      + \frac{4}{81}\frac{\tau_i^4}{16}
        \begin{tikzpicture}[baseline={([yshift=-.5ex]current bounding box.center)},scale=1]
          \useasboundingbox (-1,-1) rectangle (1,1);
          \coordinate (v1) at (30:0.77);
          \coordinate (v2) at (150:0.77);
          \coordinate (v3) at (210:0.77);
          \coordinate (v4) at (330:0.77);
          \draw[G] (v1) -- (v2);
          \draw[G] (v3) -- (v4);
          \draw[G] (v4) arc (-30:30:0.77);
          \draw[H] (v1) arc (30:150:0.77);
          \draw[G] (v2) arc (150:210:0.77);
          \draw[H] (v3) arc (210:330:0.77);
          \draw (0,0) node {\textcolor{red}{$i$}};
        \end{tikzpicture}
      + \textcolor{red}{\frac{1}{81}} \frac{\tau_i^2 \tau_j^2}{16}
        \begin{tikzpicture}[baseline={([yshift=-.5ex]current bounding box.center)},scale=1]
          \useasboundingbox (-1,-1) rectangle (1,1);
          \coordinate (v1) at (30:0.77);
          \coordinate (v2) at (150:0.77);
          \coordinate (v3) at (210:0.77);
          \coordinate (v4) at (330:0.77);
          \draw[G] (v1) -- (v2);
          \draw[G] (v3) -- (v4);
          \draw[H] (v4) arc (-30:30:0.77);
          \draw[G] (v1) arc (30:150:0.77);
          \draw[H] (v2) arc (150:210:0.77);
          \draw[G] (v3) arc (210:330:0.77);
          \draw (0,0.55) node {\textcolor{red}{$i$}};
          \draw (0,-0.55) node {\textcolor{red}{$j$}};         
        \end{tikzpicture}
        \label{eq:veff3i5}
  \\
  %
  %
      & + \frac{\tau_{0}^4}{24}
        \begin{tikzpicture}[baseline={([yshift=-.5ex]current bounding box.center)},scale=1]
          \useasboundingbox (-1,-1) rectangle (1,1);
          \coordinate (v0) at (0,0);
          \coordinate (v1) at (90:0.77);
          \coordinate (v2) at (210:0.77);
          \coordinate (v3) at (330:0.77);
          \draw[H] (v0) -- (v1);
          \draw[H] (v0) -- (v2);
          \draw[H] (v0) -- (v3);
          \draw[H] (v2) arc (-150:-30:0.77);
          \draw[H] (v3) arc (-30:90:0.77);
          \draw[H] (v1) arc (90:210:0.77);          
        \end{tikzpicture}
        + \frac{4}{9}\frac{\tau_0 \tau_i^3}{24}
        \begin{tikzpicture}[baseline={([yshift=-.5ex]current bounding box.center)},scale=1]
          \useasboundingbox (-1,-1) rectangle (1,1);
          \coordinate (v0) at (0,0);
          \coordinate (v1) at (90:0.77);
          \coordinate (v2) at (210:0.77);
          \coordinate (v3) at (330:0.77);
          \draw[H] (v0) -- (v1);
          \draw[H] (v0) -- (v2);
          \draw[H] (v0) -- (v3);
          \draw[G] (v2) arc (-150:-30:0.77);
          \draw[G] (v3) arc (-30:90:0.77);
          \draw[G] (v1) arc (90:210:0.77);          
          \draw (0,-0.5) node {\textcolor{red}{$i$}};
        \end{tikzpicture}
        + \frac{1}{9}\frac{\tau_i^4}{24}
        \begin{tikzpicture}[baseline={([yshift=-.5ex]current bounding box.center)},scale=1]
          \useasboundingbox (-1,-1) rectangle (1,1);
          \coordinate (v0) at (0,0);
          \coordinate (v1) at (90:0.77);
          \coordinate (v2) at (210:0.77);
          \coordinate (v3) at (330:0.77);
          \draw[H] (v0) -- (v1);
          \draw[G] (v0) -- (v2);
          \draw[G] (v0) -- (v3);
          \draw[H] (v2) arc (-150:-30:0.77);
          \draw[G] (v3) arc (-30:90:0.77);
          \draw[G] (v1) arc (90:210:0.77);          
          \draw (-0.4,0.3) node {\textcolor{red}{$i$}};
        \end{tikzpicture}.
        \label{eq:veff3i6}
\end{align}
Here, use the following notation: in each term, the symmetry and color factors
are shown explicitly and the part corresponding to the scalar loop integral
with all divergences properly subtracted is depicted as a figure.
We use loop functions with subtracted divergences as introduced in
Ref.~\cite{Martin:2017lqn} to separate the calculation of the complicated
three-loop integrals from the renormalization and the calculation of the
simple lower-loop integrals.

\section{Evaluation of the three-loop integrals}
\label{sec:details}

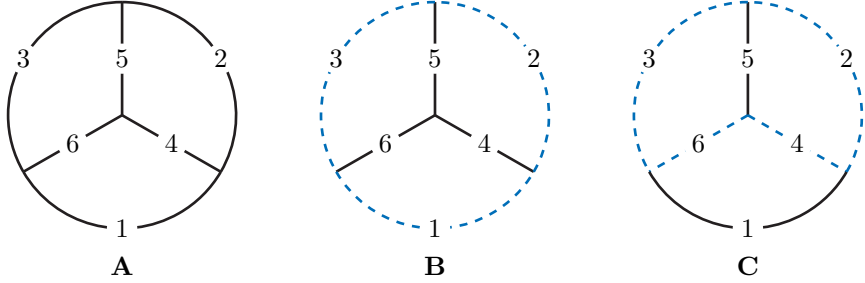
\begin{figure}[h]
  \centering
  \begin{tikzpicture}[baseline={([yshift=-.5ex]current bounding box.center)},scale=1]
    \useasboundingbox (-2,-2) rectangle (2,2);
    \coordinate (v0) at (0,0);
    \coordinate (v1) at (90:1.5);
    \coordinate (v2) at (210:1.5);
    \coordinate (v3) at (330:1.5);
    \draw[H] (v0) -- (v1);
    \draw[H] (v0) -- (v2);
    \draw[H] (v0) -- (v3);
    \draw[H] (v2) arc (-150:-30:1.5);
    \draw[H] (v3) arc (-30:90:1.5);
    \draw[H] (v1) arc (90:210:1.5);
    \fill[fill=white] (-90:1.5) circle [radius=0.2] node {$1$};
    \fill[fill=white] (30:1.5) circle [radius=0.2] node {$2$};
    \fill[fill=white] (150:1.5) circle [radius=0.2] node {$3$};
    \fill[fill=white] (-30:0.75) circle [radius=0.2] node {$4$};
    \fill[fill=white] (90:0.75) circle [radius=0.2] node {$5$};
    \fill[fill=white] (210:0.75) circle [radius=0.2] node {$6$};
    \node at (0,-2) {\texttt{\bf A}};
  \end{tikzpicture}
  \begin{tikzpicture}[baseline={([yshift=-.5ex]current bounding box.center)},scale=1]
    \useasboundingbox (-2,-2) rectangle (2,2);
    \coordinate (v0) at (0,0);
    \coordinate (v1) at (90:1.5);
    \coordinate (v2) at (210:1.5);
    \coordinate (v3) at (330:1.5);
    \draw[H] (v0) -- (v1);
    \draw[H] (v0) -- (v2);
    \draw[H] (v0) -- (v3);
    \draw[G] (v2) arc (-150:-30:1.5);
    \draw[G] (v3) arc (-30:90:1.5);
    \draw[G] (v1) arc (90:210:1.5);          
    \fill[fill=white] (-90:1.5) circle [radius=0.2] node {$1$};
    \fill[fill=white] (30:1.5) circle [radius=0.2] node {$2$};
    \fill[fill=white] (150:1.5) circle [radius=0.2] node {$3$};
    \fill[fill=white] (-30:0.75) circle [radius=0.2] node {$4$};
    \fill[fill=white] (90:0.75) circle [radius=0.2] node {$5$};
    \fill[fill=white] (210:0.75) circle [radius=0.2] node {$6$};
    \node at (0,-2) {\texttt{\bf B}};
  \end{tikzpicture}
  \begin{tikzpicture}[baseline={([yshift=-.5ex]current bounding box.center)},scale=1]
    \useasboundingbox (-2,-2) rectangle (2,2);
    \coordinate (v0) at (0,0);
    \coordinate (v1) at (90:1.5);
    \coordinate (v2) at (210:1.5);
    \coordinate (v3) at (330:1.5);
    \draw[H] (v0) -- (v1);
    \draw[G] (v0) -- (v2);
    \draw[G] (v0) -- (v3);
    \draw[H] (v2) arc (-150:-30:1.5);
    \draw[G] (v3) arc (-30:90:1.5);
    \draw[G] (v1) arc (90:210:1.5);          
    \fill[fill=white] (-90:1.5) circle [radius=0.2] node {$1$};
    \fill[fill=white] (30:1.5) circle [radius=0.2] node {$2$};
    \fill[fill=white] (150:1.5) circle [radius=0.2] node {$3$};
    \fill[fill=white] (-30:0.75) circle [radius=0.2] node {$4$};
    \fill[fill=white] (90:0.75) circle [radius=0.2] node {$5$};
    \fill[fill=white] (210:0.75) circle [radius=0.2] node {$6$};
    \node at (0,-2) {\texttt{\bf C}};
  \end{tikzpicture}
  \caption{Topologies of the master integrals $J_A$, $J_B$, and $J_C$.
    Solid and dashed lines have mass squared $m_2^2=1$ and $m_1^2=x m_2^2$,
    respectively.}
  \label{fig:topoABC}
\end{figure}

Let us introduce some general notations.
All the three-loop integrals which appear in this calculation can be mapped,
depending on the masses attached to the internal lines, on one of the three
topologies {\bf A}, {\bf B}, or {\bf C} shown in Fig.~\ref{fig:topoABC}.
We set $m_2=1$ and define the ratio $x=m_1^2/m_2^2$.
After this rescaling, the respective master integrals read
\begin{align}
  J^A_{n_1\dots n_6} & =\int                    \frac{d[k_1]d[k_2]d[k_3]}{(k_1^2+1)^{n_1}(k_2^2+1)^{n_2}(k_3^2+1)^{n_3}(k_{12}^2+1)^{n_4}(k_{23}^2+1)^{n_5}(k_{31}^2+1)^{n_6}}\,,\nonumber\\
  J^B_{n_1\dots n_6} & =\int                       \frac{d[k_1]d[k_2]d[k_3]}{(k_1^2+\ctx)^{n_1}(k_2^2+\ctx)^{n_2}(k_3^2+\ctx)^{n_3}(k_{12}^2+1)^{n_4}(k_{23}^2+1)^{n_5}(k_{31}^2+1)^{n_6}}\,,\nonumber\\
  J^C_{n_1\dots n_6} & =\int \frac{d[k_1]d[k_2]d[k_3]}{(k_1^2+1)^{n_1}(k_2^2+\ctx)^{n_2}(k_3^2+\ctx)^{n_3}(k_{12}^2+\ctx)^{n_4}(k_{23}^2+1)^{n_5}(k_{31}^2+\ctx)^{n_6}} \,,
\end{align}
where $k_{ab}=k_a-k_b$ and the loop integration measure is defined as
$d[k_i]=d^d k/\pi^{d/2} e^{\ep\gamma_E}$, with $d=4-2\ep$ being the dimension of
space-time and $\gamma_E$ being Euler's constant.

The integrals of topology {\bf A} are single-scale integrals and have been
known for a long time \cite{Steinhauser:2000ry}.
Recently, they have been evaluated through weight six \cite{Kniehl:2017ikj}.
The integrals of topologies {\bf B} and {\bf C}, appearing in
lines~\eqref{eq:veff3i1}--\eqref{eq:veff3i6}, have two mass scales and depend
on the ratio $x$.
For each of these two topologies, we have constructed a set of reduction rules
with the help of the \texttt{LiteRed} package \cite{Lee:2012cn} and identified
the set of the master integrals.
Differentiating all the master integrals with respect to $x$ and reducing the
right-hand sides to the set of master integrals, we obtain the following
systems of differential equations:
\begin{equation}
  \label{eq:de-sysBC}
  \frac{\partial J^{\rm b}_i}{\partial x} = M^{\rm b}_{ij} J^{\rm b}_j, \qquad
  \frac{\partial J^{\rm c}_i}{\partial x} = M^{\rm c}_{ij} J^{\rm c}_j \,.
\end{equation}
The elements of the matrices $M^{\rm b}$ and $M^{\rm c}$ are rational functions of
the mass ratio $x$ and the space-time dimension $d$.
It is convenient to switch from the original basis $\{J^{\rm b},J^{\rm c}\}$ to
the new, \textit{canonical} basis $\{g^{\rm b},g^{\rm c}\}$ of master integrals
introduced in Ref.~\cite{Henn:2013pwa}.
The integrals in the canonical basis possess the property that the
coefficients of their expansions in $\ep$ have uniform transcendentality
weight, and the matrix of the differential equations has a special, so-called
$\ep$ form.
An early discussion of the uniform transcendentality weight of the $\ep$
expansion may be found in Refs.~\cite{Kotikov:2010gf,Kotikov:2012ac}.

For topologies {\bf B} and {\bf C}, we choose the following sets of master
integrals to be transformed to the canonical bases by suitable transformation
matrices:
\begin{equation}
\begin{pmatrix}
  J^{\rm b}_{0, 0, 1, 0, 1, 1} \\
  J^{\rm b}_{0, 1, 1, 0, 0, 1} \\
  J^{\rm b}_{1, 1, 1, 0, 0, 0} \\
  J^{\rm b}_{0, 0, 1, 1, 1, 1} \\
  J^{\rm b}_{1, 1, 1, 0, 0, 1} \\
  J^{\rm b}_{0, 1, 1, 0, 1, 1} \\
  J^{\rm b}_{0, 1, 1, 1, 0, 1} \\
  J^{\rm b}_{0, 2, 1, 1, 0, 1} \\
  J^{\rm b}_{1, 1, 1, 0, 1, 1} \\
  J^{\rm b}_{0, 1, 1, 1, 1, 1} \\
  J^{\rm b}_{1, 1, 1, 1, 1, 1}
\end{pmatrix}
= T_{\rm b} \cdot
\begin{pmatrix}
  g^{\rm b}_1\\
  g^{\rm b}_2\\
  g^{\rm b}_3\\
  g^{\rm b}_4\\
  g^{\rm b}_5\\
  g^{\rm b}_6\\
  g^{\rm b}_7\\
  g^{\rm b}_8\\
  g^{\rm b}_9\\
  g^{\rm b}_{10}\\
  g^{\rm b}_{11}\\
\end{pmatrix},
\quad
\begin{pmatrix}
  J^{\rm c}_{0, 0, 1, 0, 1, 1} \\
  J^{\rm c}_{0, 0, 1, 1, 0, 1} \\
  J^{\rm c}_{1, 0, 0, 0, 1, 1}\\
  J^{\rm c}_{0, 0, 1, 1, 1, 1}\\
  J^{\rm c}_{ 0, 1, 1, 1, 0, 1}\\
  J^{\rm c}_{1, 0, 0, 1, 1, 1}\\
  J^{\rm c}_{0, 1, 1, 1, 1, 1}\\
  J^{\rm c}_{ 1, 0, 1, 1, 1, 0}\\
  J^{\rm c}_{2, 0, 1, 1, 1, 0}\\
  J^{\rm c}_{1, 0, 1, 1, 1, 1}\\
  J^{\rm c}_{ 1, 1, 1, 1, 1, 1}
\end{pmatrix}
=
T_{\rm c} \cdot
\begin{pmatrix}
  g^{\rm c}_1\\
  g^{\rm c}_2\\
  g^{\rm c}_3\\
  g^{\rm c}_4\\
  g^{\rm c}_5\\
  g^{\rm c}_6\\
  g^{\rm c}_7\\
  g^{\rm c}_8\\
  g^{\rm c}_9\\
  g^{\rm c}_{10}\\
  g^{\rm c}_{11}\\
\end{pmatrix}.
\label{eq:canonical-basis}
\end{equation}

\begin{figure}[h]
  \centering
  \begin{tikzpicture}
    \useasboundingbox (-1,-2.5) rectangle (9,1);
    \draw[anchor=south] (0,0) node {$0$};

    \draw[color=red,very thick,decoration = {zigzag,segment length = 0.35cm, amplitude = 0.9mm},decorate] (0,0) -- (1.5,0);
    \draw[anchor=south] (1.5,0) node {$\frac{1}{4}$};

    \draw[color=MidnightBlue,very thick,dotted] (1.5,0) -- (2,0);
    \draw[anchor=south] (2,0) node {$\frac{1}{3}$};

    \draw[color=ForestGreen,very thick,dashed] (2,0) -- (6,0);
    \draw[anchor=south] (6,0) node {$1$};
    \draw[color=black,very thick, ->, >=stealth'] (6,0) -- (8,0);
    \draw[anchor=south] (8,0) node {$\infty$};

    \draw[color=black,thick] (0,-0.15) -- (0,0.0);
    \draw[color=black,thick] (1.5,-0.15) -- (1.5,0.0);
    \draw[color=black,thick] (2,-0.15) -- (2,0.0);
    \draw[color=black,thick] (6,-0.15) -- (6,0.0);

    \draw (4,1) node {$x$};
  \end{tikzpicture}
  \quad
  \begin{tikzpicture}
    \useasboundingbox (-2.5,-2.5) rectangle (2.5,1);
    
    \draw (-90:2) arc (-90:-180:2);
    \draw (-2.5,0) -- (0,0);
    \draw (2,0) -- (2.5,0);
    \draw (0,0.5) -- (0,-2.5);
    \draw[anchor=south east] (0,0) node {$0$};
    \draw[color=red,decoration = {zigzag,segment length = 0.35cm, amplitude = 0.9mm},decorate,very thick] (0,0) -- (2,0);
    
    \draw[anchor=south] (0:2) node {$1$};
    \draw[color=MidnightBlue,very thick,dotted] (0:2) arc (0:-30:2);
    \draw[anchor=north west] (-30:2) node {$e^{\frac{-i \pi}{6}}$};
    \draw[color=ForestGreen,very thick,dashed] (-30:2) arc (-30:-60:2);
    \draw[anchor=north west] (-60:2) node {$e^{\frac{-i \pi}{3}}$};
    \draw[color=black,very thick, ->, >=stealth'] (-60:2) arc (-60:-90:2);
    \draw[anchor=center] (0,1) node {$y$};
  \end{tikzpicture}  
  \caption{Mapping in Eq.~(\ref{eq:xtoy}).}
  \label{fig:y2x-mapping}
\end{figure}
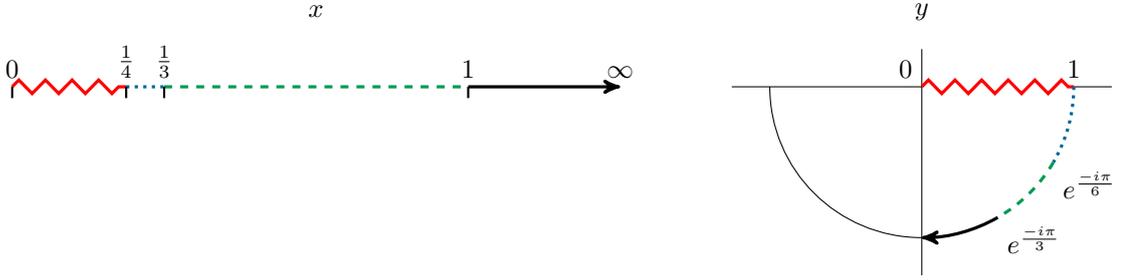
To find the explicit forms of the transformation matrices $T_{\rm b}$ and
$T_{\rm c}$ and of the matrix of the differential equations in the $\ep$ form,
we use the public tools \texttt{Fuchsia} \cite{Gituliar:2017vzm} and
\texttt{CANONICA} \cite{Meyer:2017joq}, which are implementations of the
algorithms of Refs.~\cite{Lee:2014ioa} and \cite{Meyer:2016slj}, respectively.
Both algorithms heavily rely on the polynomial, rational form of the
transformation matrices $T_{\rm b}$ and $T_{\rm c}$.
To fulfill this condition, we first make the variable transformation
\begin{equation}
  \label{eq:xtoy}
  x=\frac{y^2}{(1+y^2)^2} \,, 
\end{equation}
illustrated in Fig.~\ref{fig:y2x-mapping}, upon which we can successfully find
the transformation matrices $T_{\rm b}$ and $T_{\rm c}$.
The corresponding systems of differential equations then take the form
\begin{equation}
  \label{eq:canonical-sys-DE}
  \frac{\partial g^{\rm b}_i}{\partial y} = \ep B_{ij} g^{\rm b}_j, \qquad   \frac{\partial g^{\rm c}_i}{\partial y} = \ep C_{ij} g^{\rm c}_j \,.
\end{equation}
Now, the matrices $B$ and $C$ are independent of the space-time dimension $d$.
We can proceed further and decompose them into sums of constant matrices with
all dependence on the kinematic variable $y$ factorized out.
These decompositions read:
\begin{align}
  \label{eq:const-mat-exp}
  B(y) =  
  & \left(f^0_0 B_{0,0}+ 
    f^0_1 B_{1,0}+ 
    f^0_2 B_{2,0}+ 
    f^0_3 B_{3,0}+ 
    f^1_3 B_{3,1}+
    f^1_4 B_{4,1}+ 
    f^0_6 B_{6,0}+ 
    f^1_6 B_{6,1}+ 
    f^1_{12} B_{12,1}+
    f^3_{12} B_{12,3}
    \right),\nonumber\\
  C(y) = 
  & \left(  f^0_0 C_{0,0}+ 
    f^0_1 C_{1,0}+ 
    f^0_2 C_{2,0}+ 
    f^0_3 C_{3,0}+ 
    f^1_3 C_{3,1} \right.
    \left. + f^1_4 C_{4,1}+ 
    f^0_6 C_{6,0}+ 
    f^1_6 C_{6,1}+ 
    f^3_{8} C_{8,3}
    \right).
\end{align}
The matrices $B_{a,b}$ and $C_{a,b}$ are constant, with rational entries, and all
dependence on $y$ is contained in the functions $f^b_a(y)$, defined as
\begin{equation}
  \label{eq:f-def}
  f^0_0(y) = \frac{1}{y},\qquad  f^b_a(y) = \frac{y^b}{\Phi_a(y)},
\end{equation}
where $\Phi_n(y)$ is the $n$-th cyclotomic polynomial relevant for our
calculation.
These are given by
\begin{align}
  \label{eq:cyclo-poly-1-12}
  \Phi_{1}(y) & = y-1\,,\nonumber\\
  \Phi_{2}(y) & = y + 1\,,\nonumber\\
  \Phi_{3}(y) & = y^2 + y + 1\,,\nonumber\\
  \Phi_{4}(y) & = y^2 + 1\,,\nonumber\\
  \Phi_{6}(y) & = y^2 - y + 1\,,\nonumber\\
  \Phi_{8}(y) & = y^4 + 1\,,\nonumber\\
  \Phi_{12}(y) & = y^4 - y^2 + 1\,.
\end{align}
As a nice property, these functions have transparent integration rules, which
lead to a special type of functions, namely iterated integrals which
generalize harmonic polylogarithms.
These so-called cyclotomic harmonic polylogarithms were introduced in
Ref.~\cite{Ablinger:2011te}.

The systems of differential equations in $\ep$ form in
Eq.~(\ref{eq:canonical-sys-DE}) have the nice property that,
after the expansion of all the master integrals in $\ep$, the differential
equations for the series coefficients completely decouple and can be written in
the form
\begin{equation}
  \label{eq:ep-exp-integ}
  g^{\rm b}_i\{\ep^n\}(y) = \int\limits_{0}^{1} dy  B_{ij} g^{\rm b}_j\{\ep^{n-1}\}(y) + \mathcal{C}_{i,n} \,,
\end{equation}
and similarly for topology {\bf C}, where $g^{\rm b}\{\ep^n\}(y)$ denotes the
$O(\ep^n)$ coefficient of the $\ep$ expansion of $g^{\rm b}(y)$.
Integration of the decompositions in Eq.~\eqref{eq:const-mat-exp} leads to an
iterative integration of the functions in Eq.~\eqref{eq:f-def}.
Starting from the lowest order of the expansion in $\ep$, which is just a
constant, we can integrate order by order introducing the definition of the
cyclotomic harmonic polylogarithm of weight zero,
$\mathcal{H}\left[ \f{}{} \right](y) = 1$, and using the rule
\begin{equation}
  \label{eq:chpl-int}
  \mathcal{H}\left[ \f{b}{a}; \f{b_1}{a_1}; \ldots ;\f{b_k}{a_k}\right](y) = \int d y f^b_a(y) \mathcal{H}\left[ \f{b_1}{a_1}; \ldots ;\f{b_k}{a_k}\right](y).
\end{equation}
At each step of integration, we need to fix the integration constant.
This is done by using the expansions of the integrals in the small-$x$ limit.
These expansions are not naive, but contain, on top of the power-like terms,
also logarithms due to subgraphs.

\section{Results}
\label{sec:results}

The results for the integrals in the canonical basis up to transcendental
weight four can be found in the ancillary files of our submission to the arXiv.
To test the validity of the obtained results, we perform a number of
comparisons with already known results and carry out numerous expansions in
different limits.
Two-scale integrals with topologies $\textbf{B}$ and $\textbf{C}$ reduce in the
limit $x\to e^{\frac{i\pi}{3}}$ to one-scale integrals with topology $\textbf{A}$
and can be calculated with the help of the \texttt{MATAD} package.
The four-line \cite{Bekavac:2009gz} and five-line \cite{Martin:2016bgz}
integrals have already been known before.
Therefore, we only present here the expressions for the most complicated
integrals with six lines.


After calculating all the needed integrals, we are ready to present our final
results for the loop functions in lines~\eqref{eq:veff3i1}--\eqref{eq:veff3i6}.
We do this for each diagram separately.
Defining $L_g=\ln(m_g^2/\mu^2)$ and $L_h=\ln(m_h^2/\mu^2)$, we have for the
diagrams with three one-loop subgraphs:
  \begin{align}
    \label{eq:dias-to-ints-AxAxA}
    \mathrm{Dia[Eq.~\ref{eq:veff3i1},1]} & = (1 - L_h)^2 L_h m_h^4\,, &&\nonumber\\
    \mathrm{Dia[Eq.~\ref{eq:veff3i1},2]} & = (1 - L_g) (1 - L_h) L_h m_g^2 m_h^2\,, &&\nonumber\\
    \mathrm{Dia[Eq.~\ref{eq:veff3i1},3]} & = L_g (1 - L_h)^2 m_h^4\,, &&\nonumber\\
    \mathrm{Dia[Eq.~\ref{eq:veff3i1},4]} & = (1 - L_g)^2 L_h m_g^4\,, &&\nonumber\\
    \mathrm{Dia[Eq.~\ref{eq:veff3i1},5]} & = (1 - L_g) L_g (1 - L_h) m_g^2 m_h^2\,, &&\nonumber\\
    \mathrm{Dia[Eq.~\ref{eq:veff3i1},6]} & = (1 - L_g)^2 L_g m_g^4\,. &&
  \end{align}
For the diagrams with convolutions of two-loop and one-loop subgraphs, we have:
  \begin{align}
    \label{eq:dias-to-ints-AxT1}
    \mathrm{Dia[Eq.~\ref{eq:veff3i2},1]} & =
                                          m_h^2 \left(
                                          -\frac{3}{2}L_h^2 + \frac{1}{2} L_h^3 + \frac{3}{2} L_h
                                          \left(1 - 3 \texttt{S2}\right) - \frac{1}{2} \left(1 - 9 \texttt{S2}\right)
                                          \right)\,,
    \nonumber\\
    \mathrm{Dia[Eq.~\ref{eq:veff3i2},2]} & = m_g^2 \left((1 - L_g) L_h - \frac{1}{2}(1 - L_g) L_h^2 + \frac{1}{2} L_g (1 - 9 \texttt{S2}) -
    \frac{1}{2} (1 - 9 \texttt{S2})\right)\,, 
    \nonumber\\
    \mathrm{Dia[Eq.~\ref{eq:veff3i2},3]} & =
    \frac{(1 - L_h) m_h^2}{8(m_h^2 -4 m_g^2)}
                                          \left(
                                          m_g^2 \left(-72 + 32 L_g + 32 L_h - 16 L_g L_h + 32 L_h^2 - 2 \pi^2\right) \right.\nonumber\\
                                        & \left. + 
                                          m_h^2 \left(-48 + 8 L_g + 48 L_h - 8 L_g L_h + 4 L_h^2 - \pi^2 - 
                                          24 L_h J^b_{0 1 1 0 1 1} \left\{1/\ep^2\right\} + 8 J^b_{0 1 1 0 1 1} \left\{1/\ep\right\}\right)
                                          \right)\,,
    \nonumber\\
    \mathrm{Dia[Eq.~\ref{eq:veff3i2},4]} & =
    -\frac{(1 - L_g) m_g^2}{8(m_h^2 - 4 m_g^2)} 
    \left(
                                          m_g^2 \left(72 - 32 L_g - 32 L_h + 16 L_g L_h - 32 L_h^2 + 2 \pi^2\right) \right.\nonumber\\
                                        & \left. + 
                                          m_h^2 \left(48 - 8 L_g - 48 L_h + 8 L_g L_h - 4 L_h^2 + \pi^2 + 
                                          24 L_h J^b_{0 1 1 0 1 1} \left\{1/\ep^2\right\} - 8 J^b_{0 1 1 0 1 1} \left\{1/\ep\right\}\right)
                                          \right)\,,
    \nonumber\\
    \mathrm{Dia[Eq.~\ref{eq:veff3i2},5]} & =
                                          \frac{1 - L_h}{8 ( m_h^2 - 4 m_g^2) } 
                                          \left(
                                          \left(16 - 64 L_g + 8 L_g^2 + 32 L_h + 32 L_g L_h - 24 L_h^2\right) m_g^2 m_h^2 \right. \nonumber\\
                                        & + 
                                          m_g^4 \left(-176 + 128 L_g - 32 L_g^2 + 64 L_h - 32 L_g L_h + 64 L_h^2 - 
                                          4 \pi^2\right) \nonumber\\
                                        & \left. + m_h^4 (40 - 40 L_h - 4 L_h^2 + \pi^2) -
                                          8 m_h^2 \left(2 m_g^2 - m_h^2\right)\left(
                                          3 L_h  J^b_{0 1 1 0 1 1} \left\{1/\ep^2\right\} - 
                                           J^b_{0 1 1 0 1 1} \left\{1/\ep\right\}\right)
                                          \right)\,,
    \nonumber\\
    \mathrm{Dia[Eq.~\ref{eq:veff3i2},6]} & =
                                          -\frac{(1 - L_g) m_g^2}{8 (m_h^2 - 4 m_g^2 ) m_h^2 } 
                                          \left(  
                                          \left(-16 + 64 L_g - 8 L_g^2 - 32 L_h - 32 L_g L_h + 24 L_h^2\right) m_g^2 m_h^2\right. \nonumber\\
                                        & 
                                          + m_g^4 \left(176 - 128 L_g + 32 L_g^2 - 64 L_h + 32 L_g L_h - 64 L_h^2 + 
                                          4 \pi^2\right) \nonumber\\
                                        & \left. + m_h^4 \left(-40 + 40 L_h + 4 L_h^2 - \pi^2\right)
                                          + 8 m_h^2 \left(2 m_g^2 - m_h^2\right)\left(3 L_h  J^{b}_{0 1 1 0 1 1} \left\{1/\ep^{2}\right\} - 
                                          J^{b}_{0 1 1 0 1 1} \left\{1/\ep\right\}\right)
                                          \right)\,.
  \end{align}
For the three-loop four-line diagrams, we have:
  \begin{align}
    \label{eq:dias-to-ints-BN}
    \mathrm{Dia[Eq.~\ref{eq:veff3i3},1]} & = \frac{m_h^4}{12} \left(1 + 210 L_h - 132 L_h^2 + 24 L_h^3 \right)\,, \nonumber\\
    \mathrm{Dia[Eq.~\ref{eq:veff3i3},2]} & = -8  m_g^2 m_h^2 (2 - 8 L_g + L_g^2 + 4 L_h + 4 L_g L_h - 3 L_h^2)
                                          - m_h^4 \left(40 - 40 L_h - 4 L_h^2 + \pi^2 \right) \nonumber\\
                                        & + 
                                          4 m_g^4 \left(44 - 32 L_g + 8 L_g^2 - 16 L_h + 8 L_g L_h - 
                                          16 L_h^2 + \pi^2 \right) \nonumber \\
                                        & + (48 L_h m_g^2 m_h^2 - 24 L_h m_h^4) J^b_{0 1 1 0 1 1} \left\{1/\ep^2\right\}
                                          + (8 m_h^4 - 16 m_g^2 m_h^2) J^b_{0 1 1 0 1 1}\left\{1/\ep\right\}\,, \nonumber\\
    \mathrm{Dia[Eq.~\ref{eq:veff3i3},3]} & = \frac{m_g^4}{12} \left(1 + 210 L_g - 132 L_g^2 + 24 L_g^3 \right)\,.
  \end{align}
For the three-loop five-line diagrams, we have:
  \begin{align}
    \label{eq:dias-to-ints-E}
    \mathrm{Dia[Eq.~\ref{eq:veff3i4},1]} & = \frac{m_h^2}{3} \left(-97 + 78 L_h - 24 L_h^2 + 3 L_h^3 + 162 \texttt{S2} - 81 L_h \texttt{S2} + 
                                          18 \zeta_3\right)\,,
    \nonumber\\
    \mathrm{Dia[Eq.~\ref{eq:veff3i4},2]} & =
                                          \frac{m_g^2}{12} \left(2 L_g^3 - 48 L_h + 24 L_g L_h - 24 L_h^2 - 12 L_g L_h^2 + 
                                          14 L_h^3 - 3 \pi^2 + 3 L_g \pi^2 + 8 \zeta_3\right) \nonumber\\
                                        & + 
                                          \frac{m_h^2}{24} \left(-24 - 528 L_h + 204 L_h^2 + 16 L_h^3 - 5 \pi^2 - 
                                          6 L_h \pi^2 + 648 L_h \texttt{S2} - 24 \texttt{T1ep} + 20 \zeta_3\right) \nonumber\\
                                        & + \frac{3}{2} L_h (2 + L_h) m_h^2 J^b_{0 1 1 0 1 1} \left\{1/\ep^2\right\}
                                          - (1 + 2 L_h) m_h^2 J^b_{0 1 1 0 1 1} \left\{1/\ep\right\} + m_h^2 J^b_{0 1 1 0 1 1} \left\{\ep^0\right\} \nonumber\\
                                        & + 
                                          \frac{9}{2} L_h^2 m_h^2 J^b_{0 1 1 1 1 1} \left\{1/\ep^2\right\} - 
                                          3 L_h m_h^2 J^b_{0 1 1 1 1 1} \left\{1/\ep\right\} + m_h^2 J^b_{0 1 1 1 1 1} \left\{\ep^0\right\}\,,
    \nonumber\\
    \mathrm{Dia[Eq.~\ref{eq:veff3i4},3]} & = \frac{m_h^2}{12} \left(-48 L_h + 4 L_h^3 - 3 \pi^2 + 3 L_h \pi^2 + 8 \zeta_3\right) +
                                          \frac{m_g^2}{12} \left(-12 + 12 L_g - 6 L_g^2 + 4 L_g^3 \right.\nonumber\\
                                        &  \left. - 96 L_h + 48 L_g L_h - 
                                          48 L_h^2 - 24 L_g L_h^2 + 28 L_h^3 - 7 \pi^2 + 6 L_g \pi^2 + 
                                          16 \zeta_3\right) \nonumber\\
                                        & + 3 L_h (2 + L_h) m_h^2 J^b_{0 1 1 0 1 1} \left\{1/\ep^2\right\}
                                          - 2 (1 + 2 L_h) m_h^2 J^b_{0 1 1 0 1 1} \left\{1/\ep\right\}
                                          + 2 m_h^2 J^b_{0 1 1 0 1 1} \left\{\ep^0\right\} \nonumber\\
                                          & + \frac{9}{2} L_h^2 m_h^2 J^b_{1 1 1 0 1 1} \left\{1/\ep^2\right\}
                                            - 3 L_h m_h^2 J^b_{1 1 1 0 1 1} \left\{1/\ep\right\}
                                            + m_h^2 J^b_{1 1 1 0 1 1} \left\{\ep^0\right\}\,,
    \nonumber\\
    \mathrm{Dia[Eq.~\ref{eq:veff3i4},4]} & = \frac{m_g^2}{6} \left(2 L_g^3 - 48 L_h + 24 L_g L_h - 24 L_h^2 - 12 L_g L_h^2 + 
                                          14 L_h^3 - 3 \pi^2 + 3 L_g \pi^2 + 8 \zeta_3 \right) \nonumber\\
                                        & + 1/12 m_h^2 \left(-12 - 36 L_h - 6 L_h^2 + 4 L_h^3 - 4 \pi^2 + 3 L_h \pi^2 + 8 \zeta_3\right) \nonumber\\
                                        & + 3 L_h (2 + L_h) m_h^2 J^b_{0 1 1 0 1 1} \left\{1/\ep^2\right\}
                                          - 2 (1 + 2 L_h) m_h^2 J^b_{0 1 1 0 1 1} \left\{1/\ep\right\}
                                          + 2 m_h^2 J^b_{0 1 1 0 1 1} \left\{\ep^0\right\} \nonumber\\
                                        &  + 9/2 L_h^2 m_h^2 J^c_{0 1 1 1 1 1} \left\{1/\ep^2\right\}
                                          - 3 L_h m_h^2 J^c_{0 1 1 1 1 1} \left\{1/\ep\right\}
                                          + m_h^2 J^c_{0 1 1 1 1 1} \left\{\ep^0\right\}\,.
  \end{align}
For the three-loop six-line diagrams of ladder type, we have:
  \begin{align}
    \label{eq:dias-to-ints-LA}
    \mathrm{Dia[Eq.~\ref{eq:veff3i5},1]} & = \frac{1}{3} \left(5 - 3 L_h^2 + L_h^3 - 27(1+L_h) \texttt{S2} - 6 \zeta_3\right)\,, &&\nonumber\\
    \mathrm{Dia[Eq.~\ref{eq:veff3i5},2]} & =
                                          -\frac{1}{36 m_h^2 (m_h^2 - 4 m_g^2)}
                                          \left(
                                          m_g^2 m_h^2 (-80 + 22 L_g^3 - 24 L_g^2 (1 - L_h) + 17 \pi^2 \right.\nonumber\\
                                        & - 3 L_g (-28 + 16 (5 - L_h) L_h + \pi^2 + 324 \texttt{S2})
                                          + 2 L_h (L_h (78 - 23 L_h) + 3 (68 - \pi^2 - 108 \texttt{S2})) \nonumber\\
                                        & + 24 \texttt{T1ep} - 
                                          36 \zeta_3) + 
                                          m_g^4 (88 - 32 L_g^3 + 96 L_g^2 (1 - 2 L_h) + 22 \pi^2 - 
                                          12 L_g (12 - 2 L_h (34 - 7 L_h) + \pi^2) \nonumber\\
                                        & + 
                                          8 L_h (L_h (30 + 49 L_h) - 3 (48 + \pi^2)) - 24 \zeta_3) + 
                                          m_h^4 (-362 + 4 L_g^3 + 12 L_g^2 (-3 + 2 L_h) - 9 \pi^2 \nonumber\\
                                        & + 
                                          6 L_g (28 + 8 (-3 + L_h) L_h + \pi^2) - 
                                          2 L_h (L_h (135 + 44 L_h) - 6 (56 + \pi^2 - 27 \texttt{S2})) + 486 \texttt{S2} \nonumber\\
                                        & + 12 \texttt{T1ep} + 6 \zeta_3) + 
                                          18 L_h m_h^2 ((4 - 18 L_h) m_g^2 + (-4 + 3 L_h) m_h^2) J^b_{0 1 1 0 1 1} \left\{1/\ep^2\right\}  \nonumber\\
                                        & + 
                                          24 m_h^2 ((-1 + 6 L_h) m_g^2 + m_h^2) J^b_{0 1 1 0 1 1} \left\{1/\ep\right\} - 
                                          12 (2 m_g^2 m_h^2 + m_h^4) J^b_{0 1 1 0 1 1} \left\{\ep^0\right\}  \nonumber\\
                                        & - 
                                          54 L_h^2 m_h^2 (2 m_g^2 + m_h^2) J^b_{0 1 1 1 1 1} \left\{1/\ep^2\right\} + 
                                          36 L_h m_h^2 (2 m_g^2 + m_h^2) J^b_{0 1 1 1 1 1} \left\{1/\ep\right\} \nonumber\\
                                        & - 
                                          12 (2 m_g^2 m_h^2 + m_h^4) J^b_{0 1 1 1 1 1} \left\{\ep^0\right\} + 
                                          108 L_h^2 (m_g - m_h) m_h^2 (m_g + m_h) J^b_{0 2 1 1 0 1} \left\{1/\ep^2\right\}\nonumber\\
                                        & \left. - 
                                          72 L_h m_h^2 (m_g^2 - m_h^2) J^b_{0 2 1 1 0 1} \left\{1/\ep\right\} + 
                                          24 (m_g^2 - m_h^2) m_h^2  J^b_{0 2 1 1 0 1} \left\{\ep^0\right\}
                                          \right)\,,
    \nonumber\\
    \mathrm{Dia[Eq.~\ref{eq:veff3i5},3]} & =
                                          \frac{1}{12 m_g^2 (m_h^2 - 4 m_g^2)}
                                          (
                                          m_g^2 m_h^2 (64 - 28 L_g^3 - 144 L_h^2 + 20 L_h^3 - 6 L_g^2 (-7 + 4 L_h) - 
                                          5 \pi^2 \nonumber\\
                                          & + 12 L_h (-4 + \pi^2) - 
                                          3 L_g (-84 + 48 L_h - 12 L_h^2 + \pi^2) + 8 \zeta_3) - 
                                            2 m_h^4 (4 + 2 L_g^3 - 54 L_h^2 \nonumber\\
                                        & - 38 L_h^3 + 6 L_g^2 (-3 + 2 L_h) - 
                                          8 \pi^2 + 6 L_h (10 + \pi^2) + 
                                          3 L_g (28 - 24 L_h + 8 L_h^2 + \pi^2) + 12 \zeta_3) \nonumber\\
                                        & + 
                                          2 m_g^4 (-68 + 30 L_g^3 - 48 L_h^2 - 86 L_h^3 + 6 L_g^2 (-17 + 12 L_h) - 
                                          15 \pi^2 + 6 L_h (44 + \pi^2) \nonumber\\
                                        & + 
                                          12 L_g (14 - 20 L_h - 2 L_h^2 + \pi^2) + 
                                          20 \zeta_3)
                                          + (72 L_h (L_g + 3 L_h) m_g^2 m_h^2 - 108 L_h^2 m_h^4) J^b_{0 
                                          1 1 0 1 1} \left\{1/\ep^2\right\} \nonumber\\
                                        & + (-24 (L_g + 6 L_h) m_g^2 m_h^2 + 72 L_h m_h^4) J^b_{0 1 1 0 1 1} \left\{1/\ep\right\}
                                          + (48 m_g^2 m_h^2 - 24 m_h^4) J^b_{0 1 1 0 1 1} \left\{\ep^0\right\} \nonumber\\
                                        & + (-108 L_h^2 m_g^2 m_h^2 + 108 L_h^2 m_h^4) J^b_{0 2 1 1 0 1} \left\{1/\ep^2\right\}
                                          + (72 L_h m_g^2 m_h^2 - 72 L_h m_h^4) J^b_{0 2 1 1 0 1} \left\{1/\ep\right\} \nonumber\\
                                        & + (-24 m_g^2 m_h^2 + 24 m_h^4) J^b_{0 2 1 1 0 1} \left\{\ep^0\right\}
                                          + (108 L_h^2 m_g^2 m_h^2 - 54 L_h^2 m_h^4) J^b_{1 1 1 0 1 1} \left\{1/\ep^2\right\} \nonumber\\
                                        & + (-72 L_h m_g^2 m_h^2 + 36 L_h m_h^4) J^b_{1 1 1 0 1 1} \left\{1/\ep\right\}
                                          + (24 m_g^2 m_h^2 - 12 m_h^4) J^b_{1 1 1 0 1 1} \left\{\ep^0\right\}
                                          )\,,
    \nonumber\\
    \mathrm{Dia[Eq.~\ref{eq:veff3i5},4]} & =
                                          \frac{1}{12  m_h^2 (m_h^2 - 4 m_g^2)}
                                          (
                                          12 L_g m_g^4 (44 + 8 L_g^2 + 8 L_g (-4 + L_h) - 
                                          16 L_h (1 + L_h) + \pi^2) \nonumber\\
                                        & + 4 m_g^2 m_h^2 (4 - 6 L_g^2 + L_g^3 + L_h^2 (12 + 7 L_h) - 3 \pi^2
                                          + 3 L_g (16 - 4 L_h (4 + L_h) + \pi^2) + 4 \zeta_3) \nonumber\\
                                        & + m_h^4 (-96 - 7 \pi^2 + L_h (60 + 6 L_h + 4 L_h^2 + 3 \pi^2) + 8 \zeta_3) + 24 m_h^4 J^b_{0 1 1 0 1 1} \left\{\ep^0\right\} \nonumber\\
                                        & + 36 L_h m_h^2 (4 L_g m_g^2 + L_h m_h^2) J^b_{0 1 1 0 1 1} \left\{1/\ep^2\right\}
                                          - 48 (L_g m_g^2 m_h^2 + L_h m_h^4) J^b_{0 1 1 0 1 1} \left\{1/\ep\right\} \nonumber\\
                                        & + 54 L_h^2 m_h^4 J^c_{0 1 1 1 1 1} \left\{1/\ep^2\right\}
                                          - 36 L_h m_h^4 J^c_{0 1 1 1 1 1} \left\{1/\ep\right\}
                                          + 12 m_h^4 J^c_{0 1 1 1 1 1} \left\{\ep^0\right\}
                                          )\,.
  \end{align}
  For the three-loop six-line diagram of Mercedes Benz type, we
  have:  
  \begin{align}
    \label{eq:dias-to-ints-BE}
    \mathrm{Dia[Eq.~\ref{eq:veff3i6},1]} & = 6 \zeta_3 L_h - \texttt{D6}\,, &&\nonumber\\
    \mathrm{Dia[Eq.~\ref{eq:veff3i6},2]} & = 6 \zeta_3 L_h - J^{\rm b}_{1 1 1 1 1 1}\left\{\ep^0\right\}\,, &&\nonumber\\
    \mathrm{Dia[Eq.~\ref{eq:veff3i6},3]} & = 6 \zeta_3 L_h - J^{\rm c}_{1 1 1 1 1 1}\left\{\ep^0\right\}\,. &&
  \end{align}
  The symbols \texttt{S2}, \texttt{T1ep}, \texttt{E3}, \texttt{DM}, \texttt{DN},
  and \texttt{D6} denote the finite and $\mathcal{O}(\ep)$ parts of
  single-scale integrals and are defined in Ref.~\cite{Steinhauser:2000ry}.
  
Here, we only present the most complicated six-line integrals:
\begin{align}
  J^b_{111111} & =
                 \texttt{DM} - 4 \pi^2 \mathcal{H}\left[\f{0}{0}; \f{4}{1}\right] +
                 (2 \pi^2 - 162 \texttt{S2}) \mathcal{H}\left[\f{0}{0}; \f{12}{1}\right] - 
                 4 \pi^2 \mathcal{H}\left[\f{0}{0}; \f{12}{3}\right] + 
                 144 \mathcal{H}\left[\f{0}{0}; \f{0}{0}; \f{0}{0}; \f{3}{0}\right] + 
                 288 \mathcal{H}\left[\f{0}{0}; \f{0}{0}; \f{0}{0}; \f{3}{1}\right] \nonumber\\
               & - 
                 384 \mathcal{H}\left[\f{0}{0}; \f{0}{0}; \f{0}{0}; \f{4}{1}\right] - 
                 144 \mathcal{H}\left[\f{0}{0}; \f{0}{0}; \f{0}{0}; \f{6}{0}\right] + 
                 288 \mathcal{H}\left[\f{0}{0}; \f{0}{0}; \f{0}{0}; \f{6}{1}\right] + 
                 96 \mathcal{H}\left[\f{0}{0}; \f{0}{0}; \f{0}{0}; \f{12}{1}\right] - 
                 192 \mathcal{H}\left[\f{0}{0}; \f{0}{0}; \f{0}{0}; \f{12}{3}\right] \nonumber\\
               & + 
                 192 \mathcal{H}\left[\f{0}{0}; \f{0}{0}; \f{3}{0}; \f{4}{1}\right] + 
                 384 \mathcal{H}\left[\f{0}{0}; \f{0}{0}; \f{3}{1}; \f{4}{1}\right] - 
                 48 \mathcal{H}\left[\f{0}{0}; \f{0}{0}; \f{4}{1}; \f{3}{0}\right] - 
                 96 \mathcal{H}\left[\f{0}{0}; \f{0}{0}; \f{4}{1}; \f{3}{1}\right] + 
                 48 \mathcal{H}\left[\f{0}{0}; \f{0}{0}; \f{4}{1}; \f{6}{0}\right] \nonumber\\
               & - 
                 96 \mathcal{H}\left[\f{0}{0}; \f{0}{0}; \f{4}{1}; \f{6}{1}\right] - 
                 192 \mathcal{H}\left[\f{0}{0}; \f{0}{0}; \f{6}{0}; \f{4}{1}\right] + 
                 384 \mathcal{H}\left[\f{0}{0}; \f{0}{0}; \f{6}{1}; \f{4}{1}\right] - 
                 48 \mathcal{H}\left[\f{0}{0}; \f{0}{0}; \f{12}{1}; \f{3}{0}\right] + 
                 48 \mathcal{H}\left[\f{0}{0}; \f{0}{0}; \f{12}{1}; \f{3}{1}\right] \nonumber\\
               & + 
                 48 \mathcal{H}\left[\f{0}{0}; \f{0}{0}; \f{12}{1}; \f{6}{0}\right] + 
                 48 \mathcal{H}\left[\f{0}{0}; \f{0}{0}; \f{12}{1}; \f{6}{1}\right] - 
                 48 \mathcal{H}\left[\f{0}{0}; \f{0}{0}; \f{12}{3}; \f{3}{0}\right] - 
                 96 \mathcal{H}\left[\f{0}{0}; \f{0}{0}; \f{12}{3}; \f{3}{1}\right] + 
                 48 \mathcal{H}\left[\f{0}{0}; \f{0}{0}; \f{12}{3}; \f{6}{0}\right] \nonumber\\
               & - 
                 96 \mathcal{H}\left[\f{0}{0}; \f{0}{0}; \f{12}{3}; \f{6}{1}\right] + 
                 192 \mathcal{H}\left[\f{0}{0}; \f{3}{0}; \f{4}{1}; \f{4}{1}\right] + 
                 384 \mathcal{H}\left[\f{0}{0}; \f{3}{1}; \f{4}{1}; \f{4}{1}\right] - 
                 48 \mathcal{H}\left[\f{0}{0}; \f{4}{1}; \f{0}{0}; \f{3}{0}\right] - 
                 96 \mathcal{H}\left[\f{0}{0}; \f{4}{1}; \f{0}{0}; \f{3}{1}\right] \nonumber\\
               & + 
                 384 \mathcal{H}\left[\f{0}{0}; \f{4}{1}; \f{0}{0}; \f{4}{1}\right] + 
                 48 \mathcal{H}\left[\f{0}{0}; \f{4}{1}; \f{0}{0}; \f{6}{0}\right] - 
                 96 \mathcal{H}\left[\f{0}{0}; \f{4}{1}; \f{0}{0}; \f{6}{1}\right] - 
                 96 \mathcal{H}\left[\f{0}{0}; \f{4}{1}; \f{3}{0}; \f{4}{1}\right] - 
                 192 \mathcal{H}\left[\f{0}{0}; \f{4}{1}; \f{3}{1}; \f{4}{1}\right] \nonumber\\
               & + 
                 96 \mathcal{H}\left[\f{0}{0}; \f{4}{1}; \f{6}{0}; \f{4}{1}\right] - 
                 192 \mathcal{H}\left[\f{0}{0}; \f{4}{1}; \f{6}{1}; \f{4}{1}\right] - 
                 192 \mathcal{H}\left[\f{0}{0}; \f{6}{0}; \f{4}{1}; \f{4}{1}\right] + 
                 384 \mathcal{H}\left[\f{0}{0}; \f{6}{1}; \f{4}{1}; \f{4}{1}\right] - 
                 48 \mathcal{H}\left[\f{0}{0}; \f{12}{1}; \f{0}{0}; \f{3}{0}\right] \nonumber\\
               & + 
                 48 \mathcal{H}\left[\f{0}{0}; \f{12}{1}; \f{0}{0}; \f{3}{1}\right] - 
                 96 \mathcal{H}\left[\f{0}{0}; \f{12}{1}; \f{0}{0}; \f{4}{1}\right] + 
                 48 \mathcal{H}\left[\f{0}{0}; \f{12}{1}; \f{0}{0}; \f{6}{0}\right] + 
                 48 \mathcal{H}\left[\f{0}{0}; \f{12}{1}; \f{0}{0}; \f{6}{1}\right] - 
                 96 \mathcal{H}\left[\f{0}{0}; \f{12}{1}; \f{3}{0}; \f{4}{1}\right] \nonumber\\
               & + 
                 96 \mathcal{H}\left[\f{0}{0}; \f{12}{1}; \f{3}{1}; \f{4}{1}\right] + 
                 96 \mathcal{H}\left[\f{0}{0}; \f{12}{1}; \f{6}{0}; \f{4}{1}\right] + 
                 96 \mathcal{H}\left[\f{0}{0}; \f{12}{1}; \f{6}{1}; \f{4}{1}\right] - 
                 48 \mathcal{H}\left[\f{0}{0}; \f{12}{3}; \f{0}{0}; \f{3}{0}\right] - 
                 96 \mathcal{H}\left[\f{0}{0}; \f{12}{3}; \f{0}{0}; \f{3}{1}\right] \nonumber\\
               & + 
                 192 \mathcal{H}\left[\f{0}{0}; \f{12}{3}; \f{0}{0}; \f{4}{1}\right] + 
                 48 \mathcal{H}\left[\f{0}{0}; \f{12}{3}; \f{0}{0}; \f{6}{0}\right] - 
                 96 \mathcal{H}\left[\f{0}{0}; \f{12}{3}; \f{0}{0}; \f{6}{1}\right] - 
                 96 \mathcal{H}\left[\f{0}{0}; \f{12}{3}; \f{3}{0}; \f{4}{1}\right] - 
                 192 \mathcal{H}\left[\f{0}{0}; \f{12}{3}; \f{3}{1}; \f{4}{1}\right] \nonumber\\
               & + 
                 96 \mathcal{H}\left[\f{0}{0}; \f{12}{3}; \f{6}{0}; \f{4}{1}\right] - 
                 192 \mathcal{H}\left[\f{0}{0}; \f{12}{3}; \f{6}{1}; \f{4}{1}\right] - 
                 48 \mathcal{H}\left[\f{4}{1}; \f{0}{0}; \f{0}{0}; \f{3}{0}\right] - 
                 96 \mathcal{H}\left[\f{4}{1}; \f{0}{0}; \f{0}{0}; \f{3}{1}\right] + 
                 192 \mathcal{H}\left[\f{4}{1}; \f{0}{0}; \f{0}{0}; \f{4}{1}\right] \nonumber\\
               & + 
                 48 \mathcal{H}\left[\f{4}{1}; \f{0}{0}; \f{0}{0}; \f{6}{0}\right] - 
                 96 \mathcal{H}\left[\f{4}{1}; \f{0}{0}; \f{0}{0}; \f{6}{1}\right] - 
                 96 \mathcal{H}\left[\f{4}{1}; \f{0}{0}; \f{3}{0}; \f{4}{1}\right] - 
                 192 \mathcal{H}\left[\f{4}{1}; \f{0}{0}; \f{3}{1}; \f{4}{1}\right] + 
                 384 \mathcal{H}\left[\f{4}{1}; \f{0}{0}; \f{4}{1}; \f{4}{1}\right] \nonumber\\
               & + 
                 96 \mathcal{H}\left[\f{4}{1}; \f{0}{0}; \f{6}{0}; \f{4}{1}\right] - 
                 192 \mathcal{H}\left[\f{4}{1}; \f{0}{0}; \f{6}{1}; \f{4}{1}\right] - 
                 192 \mathcal{H}\left[\f{4}{1}; \f{3}{0}; \f{4}{1}; \f{4}{1}\right] - 
                 384 \mathcal{H}\left[\f{4}{1}; \f{3}{1}; \f{4}{1}; \f{4}{1}\right] + 
                 192 \mathcal{H}\left[\f{4}{1}; \f{6}{0}; \f{4}{1}; \f{4}{1}\right] \nonumber\\
               & - 
                 384 \mathcal{H}\left[\f{4}{1}; \f{6}{1}; \f{4}{1}; \f{4}{1}\right] - 
                 48 \mathcal{H}\left[\f{12}{1}; \f{0}{0}; \f{0}{0}; \f{3}{0}\right] + 
                 48 \mathcal{H}\left[\f{12}{1}; \f{0}{0}; \f{0}{0}; \f{3}{1}\right] + 
                 48 \mathcal{H}\left[\f{12}{1}; \f{0}{0}; \f{0}{0}; \f{6}{0}\right] + 
                 48 \mathcal{H}\left[\f{12}{1}; \f{0}{0}; \f{0}{0}; \f{6}{1}\right] \nonumber\\
               & - 
                 96 \mathcal{H}\left[\f{12}{1}; \f{0}{0}; \f{3}{0}; \f{4}{1}\right] + 
                 96 \mathcal{H}\left[\f{12}{1}; \f{0}{0}; \f{3}{1}; \f{4}{1}\right] - 
                 192 \mathcal{H}\left[\f{12}{1}; \f{0}{0}; \f{4}{1}; \f{4}{1}\right] + 
                 96 \mathcal{H}\left[\f{12}{1}; \f{0}{0}; \f{6}{0}; \f{4}{1}\right] + 
                 96 \mathcal{H}\left[\f{12}{1}; \f{0}{0}; \f{6}{1}; \f{4}{1}\right] \nonumber\\
               & - 
                 192 \mathcal{H}\left[\f{12}{1}; \f{3}{0}; \f{4}{1}; \f{4}{1}\right] + 
                 192 \mathcal{H}\left[\f{12}{1}; \f{3}{1}; \f{4}{1}; \f{4}{1}\right] + 
                 192 \mathcal{H}\left[\f{12}{1}; \f{6}{0}; \f{4}{1}; \f{4}{1}\right] + 
                 192 \mathcal{H}\left[\f{12}{1}; \f{6}{1}; \f{4}{1}; \f{4}{1}\right] - 
                 48 \mathcal{H}\left[\f{12}{3}; \f{0}{0}; \f{0}{0}; \f{3}{0}\right] \nonumber\\
               & - 
                 96 \mathcal{H}\left[\f{12}{3}; \f{0}{0}; \f{0}{0}; \f{3}{1}\right] + 
                 48 \mathcal{H}\left[\f{12}{3}; \f{0}{0}; \f{0}{0}; \f{6}{0}\right] - 
                 96 \mathcal{H}\left[\f{12}{3}; \f{0}{0}; \f{0}{0}; \f{6}{1}\right] - 
                 96 \mathcal{H}\left[\f{12}{3}; \f{0}{0}; \f{3}{0}; \f{4}{1}\right] - 
                 192 \mathcal{H}\left[\f{12}{3}; \f{0}{0}; \f{3}{1}; \f{4}{1}\right] \nonumber\\
               & + 
                 384 \mathcal{H}\left[\f{12}{3}; \f{0}{0}; \f{4}{1}; \f{4}{1}\right] + 
                 96 \mathcal{H}\left[\f{12}{3}; \f{0}{0}; \f{6}{0}; \f{4}{1}\right] - 
                 192 \mathcal{H}\left[\f{12}{3}; \f{0}{0}; \f{6}{1}; \f{4}{1}\right] - 
                 192 \mathcal{H}\left[\f{12}{3}; \f{3}{0}; \f{4}{1}; \f{4}{1}\right] - 
                 384 \mathcal{H}\left[\f{12}{3}; \f{3}{1}; \f{4}{1}; \f{4}{1}\right] \nonumber\\
               & + 
                 192 \mathcal{H}\left[\f{12}{3}; \f{6}{0}; \f{4}{1}; \f{4}{1}\right] - 
                 384 \mathcal{H}\left[\f{12}{3}; \f{6}{1}; \f{4}{1}; \f{4}{1}\right] + 
                 24 \mathcal{H}\left[\f{4}{1}\right] \zeta_3 -
                 24 \mathcal{H}\left[\f{12}{3}\right] \zeta_3 \nonumber\\
               & + 
                 \mathcal{H}\left[\f{12}{1}\right] (143 + 6 \texttt{E3} + 6 \pi^2 - 162 \texttt{S2} - 6 \texttt{T1ep} + 16 \zeta_3) \nonumber
  \\
               & -2 \log y (
                 \pi^2 \mathcal{H}\left[\f{12}{1}\right] -
                 81 \texttt{S2} \mathcal{H}\left[\f{12}{1}\right] - 
                 2 \pi^2 \mathcal{H}\left[\f{12}{3}\right] + 
                 48 \mathcal{H}\left[\f{0}{0}; \f{0}{0}; \f{3}{0}\right] + 
                 96 \mathcal{H}\left[\f{0}{0}; \f{0}{0}; \f{3}{1}\right] - 
                 144 \mathcal{H}\left[\f{0}{0}; \f{0}{0}; \f{4}{1}\right] \nonumber\\
               & - 
                 48 \mathcal{H}\left[\f{0}{0}; \f{0}{0}; \f{6}{0}\right] + 
                 96 \mathcal{H}\left[\f{0}{0}; \f{0}{0}; \f{6}{1}\right] + 
                 48 \mathcal{H}\left[\f{0}{0}; \f{0}{0}; \f{12}{1}\right] - 
                 96 \mathcal{H}\left[\f{0}{0}; \f{0}{0}; \f{12}{3}\right] + 
                 48 \mathcal{H}\left[\f{0}{0}; \f{3}{0}; \f{4}{1}\right] + 
                 96 \mathcal{H}\left[\f{0}{0}; \f{3}{1}; \f{4}{1}\right] \nonumber\\
               & - 
                 24 \mathcal{H}\left[\f{0}{0}; \f{4}{1}; \f{3}{0}\right] - 
                 48 \mathcal{H}\left[\f{0}{0}; \f{4}{1}; \f{3}{1}\right] + 
                 24 \mathcal{H}\left[\f{0}{0}; \f{4}{1}; \f{6}{0}\right] - 
                 48 \mathcal{H}\left[\f{0}{0}; \f{4}{1}; \f{6}{1}\right] - 
                 48 \mathcal{H}\left[\f{0}{0}; \f{6}{0}; \f{4}{1}\right] + 
                 96 \mathcal{H}\left[\f{0}{0}; \f{6}{1}; \f{4}{1}\right] \nonumber\\
               & - 
                 24 \mathcal{H}\left[\f{0}{0}; \f{12}{1}; \f{3}{0}\right] + 
                 24 \mathcal{H}\left[\f{0}{0}; \f{12}{1}; \f{3}{1}\right] + 
                 24 \mathcal{H}\left[\f{0}{0}; \f{12}{1}; \f{6}{0}\right] + 
                 24 \mathcal{H}\left[\f{0}{0}; \f{12}{1}; \f{6}{1}\right] - 
                 24 \mathcal{H}\left[\f{0}{0}; \f{12}{3}; \f{3}{0}\right] - 
                 48 \mathcal{H}\left[\f{0}{0}; \f{12}{3}; \f{3}{1}\right] \nonumber\\
               & + 
                 24 \mathcal{H}\left[\f{0}{0}; \f{12}{3}; \f{6}{0}\right] - 
                 48 \mathcal{H}\left[\f{0}{0}; \f{12}{3}; \f{6}{1}\right] - 
                 24 \mathcal{H}\left[\f{4}{1}; \f{0}{0}; \f{3}{0}\right] - 
                 48 \mathcal{H}\left[\f{4}{1}; \f{0}{0}; \f{3}{1}\right] + 
                 96 \mathcal{H}\left[\f{4}{1}; \f{0}{0}; \f{4}{1}\right] + 
                 24 \mathcal{H}\left[\f{4}{1}; \f{0}{0}; \f{6}{0}\right] \nonumber\\
               & - 
                 48 \mathcal{H}\left[\f{4}{1}; \f{0}{0}; \f{6}{1}\right] - 
                 48 \mathcal{H}\left[\f{4}{1}; \f{3}{0}; \f{4}{1}\right] - 
                 96 \mathcal{H}\left[\f{4}{1}; \f{3}{1}; \f{4}{1}\right] + 
                 48 \mathcal{H}\left[\f{4}{1}; \f{6}{0}; \f{4}{1}\right] - 
                 96 \mathcal{H}\left[\f{4}{1}; \f{6}{1}; \f{4}{1}\right] - 
                 24 \mathcal{H}\left[\f{12}{1}; \f{0}{0}; \f{3}{0}\right] \nonumber\\
               & + 
                 24 \mathcal{H}\left[\f{12}{1}; \f{0}{0}; \f{3}{1}\right] - 
                 48 \mathcal{H}\left[\f{12}{1}; \f{0}{0}; \f{4}{1}\right] + 
                 24 \mathcal{H}\left[\f{12}{1}; \f{0}{0}; \f{6}{0}\right] + 
                 24 \mathcal{H}\left[\f{12}{1}; \f{0}{0}; \f{6}{1}\right] - 
                 48 \mathcal{H}\left[\f{12}{1}; \f{3}{0}; \f{4}{1}\right] + 
                 48 \mathcal{H}\left[\f{12}{1}; \f{3}{1}; \f{4}{1}\right] \nonumber\\
               & + 
                 48 \mathcal{H}\left[\f{12}{1}; \f{6}{0}; \f{4}{1}\right] + 
                 48 \mathcal{H}\left[\f{12}{1}; \f{6}{1}; \f{4}{1}\right] - 
                 24 \mathcal{H}\left[\f{12}{3}; \f{0}{0}; \f{3}{0}\right] - 
                 48 \mathcal{H}\left[\f{12}{3}; \f{0}{0}; \f{3}{1}\right] + 
                 96 \mathcal{H}\left[\f{12}{3}; \f{0}{0}; \f{4}{1}\right] + 
                 24 \mathcal{H}\left[\f{12}{3}; \f{0}{0}; \f{6}{0}\right] \nonumber\\
               & - 
                 48 \mathcal{H}\left[\f{12}{3}; \f{0}{0}; \f{6}{1}\right] - 
                 48 \mathcal{H}\left[\f{12}{3}; \f{3}{0}; \f{4}{1}\right] - 
                 96 \mathcal{H}\left[\f{12}{3}; \f{3}{1}; \f{4}{1}\right] + 
                 48 \mathcal{H}\left[\f{12}{3}; \f{6}{0}; \f{4}{1}\right] - 
                 96 \mathcal{H}\left[\f{12}{3}; \f{6}{1}; \f{4}{1}\right]) \nonumber
  \\
               & + 24 \log^2 y
                 \left( \mathcal{H}\left[\f{0}{0}; \f{3}{0}\right] + 
                 2 \mathcal{H}\left[\f{0}{0}; \f{3}{1}\right] - 
                 4 \mathcal{H}\left[\f{0}{0}; \f{4}{1}\right] - 
                 \mathcal{H}\left[\f{0}{0}; \f{6}{0}\right] + 
                 2 \mathcal{H}\left[\f{0}{0}; \f{6}{1}\right] + 
                 2 \mathcal{H}\left[\f{0}{0}; \f{12}{1}\right] - 
                 4 \mathcal{H}\left[\f{0}{0}; \f{12}{3}\right] \right. \nonumber\\
               & - 
                 \mathcal{H}\left[\f{4}{1}; \f{3}{0}\right] -
                 2 \mathcal{H}\left[\f{4}{1}; \f{3}{1}\right] +
                 \mathcal{H}\left[\f{4}{1}; \f{6}{0}\right] - 
                 2 \mathcal{H}\left[\f{4}{1}; \f{6}{1}\right] - 
                 \mathcal{H}\left[\f{12}{1}; \f{3}{0}\right] +
                 \mathcal{H}\left[\f{12}{1}; \f{3}{1}\right] +
                 \mathcal{H}\left[\f{12}{1}; \f{6}{0}\right] + 
                 \mathcal{H}\left[\f{12}{1}; \f{6}{1}\right] \nonumber\\
               & - \left.
                 \mathcal{H}\left[\f{12}{3}; \f{3}{0}\right] - 
                 2 \mathcal{H}\left[\f{12}{3}; \f{3}{1}\right] + 
                 \mathcal{H}\left[\f{12}{3}; \f{6}{0}\right] - 
                 2 \mathcal{H}\left[\f{12}{3}; \f{6}{1}\right]\right)\,,
\nonumber\end{align}

\begin{align}
  \label{eq:jC111111}
  J^c_{111111} & =
                 \texttt{DN} + \frac{32}{3} \pi^2 \mathcal{H}\left[\f{0}{0}; \f{8}{3} \right] - 
                 192 \mathcal{H}\left[\f{0}{0}; \f{0}{0}; \f{0}{0}; \f{3}{0} \right] - 
                 384 \mathcal{H}\left[\f{0}{0}; \f{0}{0}; \f{0}{0}; \f{3}{1} \right] + 
                 384 \mathcal{H}\left[\f{0}{0}; \f{0}{0}; \f{0}{0}; \f{4}{1} \right] + 
                 192 \mathcal{H}\left[\f{0}{0}; \f{0}{0}; \f{0}{0}; \f{6}{0} \right] \nonumber\\
               & - 
                 384 \mathcal{H}\left[\f{0}{0}; \f{0}{0}; \f{0}{0}; \f{6}{1} \right] + 
                 384 \mathcal{H}\left[\f{0}{0}; \f{0}{0}; \f{0}{0}; \f{8}{3} \right] - 
                 256 \mathcal{H}\left[\f{0}{0}; \f{0}{0}; \f{3}{0}; \f{4}{1} \right] - 
                 512 \mathcal{H}\left[\f{0}{0}; \f{0}{0}; \f{3}{1}; \f{4}{1} \right] + 
                 64 \mathcal{H}\left[\f{0}{0}; \f{0}{0}; \f{4}{1}; \f{3}{0} \right] \nonumber\\
               & + 
                 128 \mathcal{H}\left[\f{0}{0}; \f{0}{0}; \f{4}{1}; \f{3}{1} \right] - 
                 64 \mathcal{H}\left[\f{0}{0}; \f{0}{0}; \f{4}{1}; \f{6}{0} \right] + 
                 128 \mathcal{H}\left[\f{0}{0}; \f{0}{0}; \f{4}{1}; \f{6}{1} \right] + 
                 256 \mathcal{H}\left[\f{0}{0}; \f{0}{0}; \f{6}{0}; \f{4}{1} \right] - 
                 512 \mathcal{H}\left[\f{0}{0}; \f{0}{0}; \f{6}{1}; \f{4}{1} \right] \nonumber\\
               & + 
                 64 \mathcal{H}\left[\f{0}{0}; \f{0}{0}; \f{8}{3}; \f{3}{0} \right] + 
                 128 \mathcal{H}\left[\f{0}{0}; \f{0}{0}; \f{8}{3}; \f{3}{1} \right] - 
                 64 \mathcal{H}\left[\f{0}{0}; \f{0}{0}; \f{8}{3}; \f{6}{0} \right] + 
                 128 \mathcal{H}\left[\f{0}{0}; \f{0}{0}; \f{8}{3}; \f{6}{1} \right] - 
                 256 \mathcal{H}\left[\f{0}{0}; \f{3}{0}; \f{4}{1}; \f{4}{1} \right] \nonumber\\
               & - 
                 512 \mathcal{H}\left[\f{0}{0}; \f{3}{1}; \f{4}{1}; \f{4}{1} \right] + 
                 64 \mathcal{H}\left[\f{0}{0}; \f{4}{1}; \f{0}{0}; \f{3}{0} \right] + 
                 128 \mathcal{H}\left[\f{0}{0}; \f{4}{1}; \f{0}{0}; \f{3}{1} \right] - 
                 512 \mathcal{H}\left[\f{0}{0}; \f{4}{1}; \f{0}{0}; \f{4}{1} \right] - 
                 64 \mathcal{H}\left[\f{0}{0}; \f{4}{1}; \f{0}{0}; \f{6}{0} \right] \nonumber\\
               & + 
                 128 \mathcal{H}\left[\f{0}{0}; \f{4}{1}; \f{0}{0}; \f{6}{1} \right] + 
                 128 \mathcal{H}\left[\f{0}{0}; \f{4}{1}; \f{3}{0}; \f{4}{1} \right] + 
                 256 \mathcal{H}\left[\f{0}{0}; \f{4}{1}; \f{3}{1}; \f{4}{1} \right] - 
                 128 \mathcal{H}\left[\f{0}{0}; \f{4}{1}; \f{6}{0}; \f{4}{1} \right] + 
                 256 \mathcal{H}\left[\f{0}{0}; \f{4}{1}; \f{6}{1}; \f{4}{1} \right] \nonumber\\
               & + 
                 256 \mathcal{H}\left[\f{0}{0}; \f{6}{0}; \f{4}{1}; \f{4}{1} \right] - 
                 512 \mathcal{H}\left[\f{0}{0}; \f{6}{1}; \f{4}{1}; \f{4}{1} \right] + 
                 64 \mathcal{H}\left[\f{0}{0}; \f{8}{3}; \f{0}{0}; \f{3}{0} \right] + 
                 128 \mathcal{H}\left[\f{0}{0}; \f{8}{3}; \f{0}{0}; \f{3}{1} \right] - 
                 256 \mathcal{H}\left[\f{0}{0}; \f{8}{3}; \f{0}{0}; \f{4}{1} \right] \nonumber\\
               & - 
                 64 \mathcal{H}\left[\f{0}{0}; \f{8}{3}; \f{0}{0}; \f{6}{0} \right] + 
                 128 \mathcal{H}\left[\f{0}{0}; \f{8}{3}; \f{0}{0}; \f{6}{1} \right] + 
                 128 \mathcal{H}\left[\f{0}{0}; \f{8}{3}; \f{3}{0}; \f{4}{1} \right] + 
                 256 \mathcal{H}\left[\f{0}{0}; \f{8}{3}; \f{3}{1}; \f{4}{1} \right] - 
                 128 \mathcal{H}\left[\f{0}{0}; \f{8}{3}; \f{6}{0}; \f{4}{1} \right] \nonumber\\
               & + 
                 256 \mathcal{H}\left[\f{0}{0}; \f{8}{3}; \f{6}{1}; \f{4}{1} \right] + 
                 64 \mathcal{H}\left[\f{4}{1}; \f{0}{0}; \f{0}{0}; \f{3}{0} \right] + 
                 128 \mathcal{H}\left[\f{4}{1}; \f{0}{0}; \f{0}{0}; \f{3}{1} \right] - 
                 512 \mathcal{H}\left[\f{4}{1}; \f{0}{0}; \f{0}{0}; \f{4}{1} \right] - 
                 64 \mathcal{H}\left[\f{4}{1}; \f{0}{0}; \f{0}{0}; \f{6}{0} \right] \nonumber\\
               & + 
                 128 \mathcal{H}\left[\f{4}{1}; \f{0}{0}; \f{0}{0}; \f{6}{1} \right] + 
                 128 \mathcal{H}\left[\f{4}{1}; \f{0}{0}; \f{3}{0}; \f{4}{1} \right] + 
                 256 \mathcal{H}\left[\f{4}{1}; \f{0}{0}; \f{3}{1}; \f{4}{1} \right] - 
                 512 \mathcal{H}\left[\f{4}{1}; \f{0}{0}; \f{4}{1}; \f{4}{1} \right] - 
                 128 \mathcal{H}\left[\f{4}{1}; \f{0}{0}; \f{6}{0}; \f{4}{1} \right] \nonumber\\
               & + 
                 256 \mathcal{H}\left[\f{4}{1}; \f{0}{0}; \f{6}{1}; \f{4}{1} \right] + 
                 256 \mathcal{H}\left[\f{4}{1}; \f{3}{0}; \f{4}{1}; \f{4}{1} \right] + 
                 512 \mathcal{H}\left[\f{4}{1}; \f{3}{1}; \f{4}{1}; \f{4}{1} \right] - 
                 256 \mathcal{H}\left[\f{4}{1}; \f{6}{0}; \f{4}{1}; \f{4}{1} \right] + 
                 512 \mathcal{H}\left[\f{4}{1}; \f{6}{1}; \f{4}{1}; \f{4}{1} \right] \nonumber\\
               & + 
                 64 \mathcal{H}\left[\f{8}{3}; \f{0}{0}; \f{0}{0}; \f{3}{0} \right] + 
                 128 \mathcal{H}\left[\f{8}{3}; \f{0}{0}; \f{0}{0}; \f{3}{1} \right] + 
                 256 \mathcal{H}\left[\f{8}{3}; \f{0}{0}; \f{0}{0}; \f{4}{1} \right] - 
                 64 \mathcal{H}\left[\f{8}{3}; \f{0}{0}; \f{0}{0}; \f{6}{0} \right] + 
                 128 \mathcal{H}\left[\f{8}{3}; \f{0}{0}; \f{0}{0}; \f{6}{1} \right] \nonumber\\
               & + 
                 128 \mathcal{H}\left[\f{8}{3}; \f{0}{0}; \f{3}{0}; \f{4}{1} \right] + 
                 256 \mathcal{H}\left[\f{8}{3}; \f{0}{0}; \f{3}{1}; \f{4}{1} \right] - 
                 512 \mathcal{H}\left[\f{8}{3}; \f{0}{0}; \f{4}{1}; \f{4}{1} \right] - 
                 128 \mathcal{H}\left[\f{8}{3}; \f{0}{0}; \f{6}{0}; \f{4}{1} \right] + 
                 256 \mathcal{H}\left[\f{8}{3}; \f{0}{0}; \f{6}{1}; \f{4}{1} \right] \nonumber\\
               & + 
                 256 \mathcal{H}\left[\f{8}{3}; \f{3}{0}; \f{4}{1}; \f{4}{1} \right] + 
                 512 \mathcal{H}\left[\f{8}{3}; \f{3}{1}; \f{4}{1}; \f{4}{1} \right] - 
                 256 \mathcal{H}\left[\f{8}{3}; \f{6}{0}; \f{4}{1}; \f{4}{1} \right] + 
                 512 \mathcal{H}\left[\f{8}{3}; \f{6}{1}; \f{4}{1}; \f{4}{1} \right] - 
                 8 \mathcal{H}\left[\f{4}{1} \right] \zeta_3 + 8 \mathcal{H}\left[\f{8}{3} \right] \zeta_3 \nonumber
  \\
               & + \frac{16}{3} \log y (\pi^2 \mathcal{H}\left[\f{4}{1} \right] -
                 2 \pi^2 \mathcal{H}\left[\f{8}{3} \right] + 
                 24 \mathcal{H}\left[\f{0}{0}; \f{0}{0}; \f{3}{0} \right] + 
                 48 \mathcal{H}\left[\f{0}{0}; \f{0}{0}; \f{3}{1} \right] - 
                 48 \mathcal{H}\left[\f{0}{0}; \f{0}{0}; \f{4}{1} \right] - 
                 24 \mathcal{H}\left[\f{0}{0}; \f{0}{0}; \f{6}{0} \right] \nonumber\\
               & + 
                 48 \mathcal{H}\left[\f{0}{0}; \f{0}{0}; \f{6}{1} \right] - 
                 72 \mathcal{H}\left[\f{0}{0}; \f{0}{0}; \f{8}{3} \right] + 
                 24 \mathcal{H}\left[\f{0}{0}; \f{3}{0}; \f{4}{1} \right] + 
                 48 \mathcal{H}\left[\f{0}{0}; \f{3}{1}; \f{4}{1} \right] - 
                 12 \mathcal{H}\left[\f{0}{0}; \f{4}{1}; \f{3}{0} \right] - 
                 24 \mathcal{H}\left[\f{0}{0}; \f{4}{1}; \f{3}{1} \right] \nonumber\\
               & + 
                 12 \mathcal{H}\left[\f{0}{0}; \f{4}{1}; \f{6}{0} \right] - 
                 24 \mathcal{H}\left[\f{0}{0}; \f{4}{1}; \f{6}{1} \right] - 
                 24 \mathcal{H}\left[\f{0}{0}; \f{6}{0}; \f{4}{1} \right] + 
                 48 \mathcal{H}\left[\f{0}{0}; \f{6}{1}; \f{4}{1} \right] - 
                 12 \mathcal{H}\left[\f{0}{0}; \f{8}{3}; \f{3}{0} \right] - 
                 24 \mathcal{H}\left[\f{0}{0}; \f{8}{3}; \f{3}{1} \right] \nonumber\\
               & + 
                 12 \mathcal{H}\left[\f{0}{0}; \f{8}{3}; \f{6}{0} \right] - 
                 24 \mathcal{H}\left[\f{0}{0}; \f{8}{3}; \f{6}{1} \right] - 
                 12 \mathcal{H}\left[\f{4}{1}; \f{0}{0}; \f{3}{0} \right] - 
                 24 \mathcal{H}\left[\f{4}{1}; \f{0}{0}; \f{3}{1} \right] + 
                 48 \mathcal{H}\left[\f{4}{1}; \f{0}{0}; \f{4}{1} \right] + 
                 12 \mathcal{H}\left[\f{4}{1}; \f{0}{0}; \f{6}{0} \right] \nonumber\\
               & - 
                 24 \mathcal{H}\left[\f{4}{1}; \f{0}{0}; \f{6}{1} \right] - 
                 24 \mathcal{H}\left[\f{4}{1}; \f{3}{0}; \f{4}{1} \right] - 
                 48 \mathcal{H}\left[\f{4}{1}; \f{3}{1}; \f{4}{1} \right] + 
                 24 \mathcal{H}\left[\f{4}{1}; \f{6}{0}; \f{4}{1} \right] - 
                 48 \mathcal{H}\left[\f{4}{1}; \f{6}{1}; \f{4}{1} \right] - 
                 12 \mathcal{H}\left[\f{8}{3}; \f{0}{0}; \f{3}{0} \right] \nonumber\\
               & - 
                 24 \mathcal{H}\left[\f{8}{3}; \f{0}{0}; \f{3}{1} \right] + 
                 48 \mathcal{H}\left[\f{8}{3}; \f{0}{0}; \f{4}{1} \right] + 
                 12 \mathcal{H}\left[\f{8}{3}; \f{0}{0}; \f{6}{0} \right] - 
                 24 \mathcal{H}\left[\f{8}{3}; \f{0}{0}; \f{6}{1} \right] - 
                 24 \mathcal{H}\left[\f{8}{3}; \f{3}{0}; \f{4}{1} \right] - 
                 48 \mathcal{H}\left[\f{8}{3}; \f{3}{1}; \f{4}{1} \right] \nonumber\\
               & + 
                 24 \mathcal{H}\left[\f{8}{3}; \f{6}{0}; \f{4}{1} \right] - 
                 48 \mathcal{H}\left[\f{8}{3}; \f{6}{1}; \f{4}{1} \right])\nonumber
  \\                 
               & -32 \log^2 y
                 (
                 \mathcal{H}\left[\f{0}{0}; \f{3}{0} \right] + 
                 2 \mathcal{H}\left[\f{0}{0}; \f{3}{1} \right] - 
                 2 \mathcal{H}\left[\f{0}{0}; \f{4}{1} \right] - 
                 \mathcal{H}\left[\f{0}{0}; \f{6}{0} \right] + 
                 2 \mathcal{H}\left[\f{0}{0}; \f{6}{1} \right] - 
                 6 \mathcal{H}\left[\f{0}{0}; \f{8}{3} \right] - 
                 \mathcal{H}\left[\f{4}{1}; \f{3}{0} \right] \nonumber\\
               & -
                 2 \mathcal{H}\left[\f{4}{1}; \f{3}{1} \right] +
                 \mathcal{H}\left[\f{4}{1}; \f{6}{0} \right] - 
                 2 \mathcal{H}\left[\f{4}{1}; \f{6}{1} \right] - 
                 \mathcal{H}\left[\f{8}{3}; \f{3}{0} \right] -
                 2 \mathcal{H}\left[\f{8}{3}; \f{3}{1} \right] +
                 \mathcal{H}\left[\f{8}{3}; \f{6}{0} \right] -
                 2 \mathcal{H}\left[\f{8}{3}; \f{6}{1} \right]
                 ) \,.
\end{align}

At high energy scales, the field strength $\phi$ is much larger than the mass
parameter $m$ in the Lagrangian, and we can neglect the latter in
Eq.~(\ref{eq:mmphi}).
Thus, the ratio $x$ is equal to $1/3$.
This limit corresponds to the value $y=e^{-i\pi/6}$.
For this massless version of the theory in Eq.~\eqref{eq:lag}, partial results
are known \cite{Chung:1998mz,Kotikov:1998uf}.
We have checked whether the expressions of all master integrals at this point
can be expressed in terms of the basis constructed in
Ref.~\cite{Kniehl:2017ikj} or a more general basis of the sixth root of unity
\cite{Henn:2015sem}.
The result is that, at weight two, this is still possible, which has already
been found in Refs.~\cite{Chung:1998mz,Kotikov:1998uf}, while, at higher
weights, some other constants appear.
Looking at the cyclotomic polynomials in Eq.~(\ref{eq:cyclo-poly-1-12}), one
can expect that the basis constructed from the twelfth root of unity would be
appropriate.

In conclusion, we have evaluated the three-loop effective potential in the
scalar sector analytically.
The result is expressed in terms of cyclotomic polylogarithms of cyclotomies
1, 2, 3, 4, 6, 8, and 12 and up to weight four.
The general three-loop vacuum master integral with two different mass scales
apparently does not lie in the class of polylogarithmic functions, while
the particular subset relevant to the scalar sector does.

\section*{Acknowledgments}

This work was supported in part by the German Federal Ministry for Education
and Research BMBF through Grant No.\ 05H2018, by the German Research
Foundation DFG through the Collaborative Research Center No.\ SFB~676
{\it Particles, Strings and the Early Universe: the Structure of Matter and
Space-Time}.
The work of AFP was supported by the Foundation for the Advancement of
Theoretical Physics and Mathematics ``BASIS.''

\appendix

\section{Numerical evaluation of cyclotomic polylogarithms}

In this appendix, we discuss the numerical evaluation of the cyclotomic
polylogarithms relevant for our study.
Let us consider the cyclotomic polylogarithm
$\mathcal{H}\left[\f{b_w}{a_w};\dots;\f{b_1}{a_1}\right](x)$
of weight $w$.
We introduce the following short-hand notation, omitting the argument $x$:
\begin{align}
  \label{ff}
  h_w = \mathcal{H}\left[\f{b_w}{a_w};\dots;\f{b_1}{a_1}\right], \quad
  h_{w-1} = \mathcal{H}\left[\f{b_{w-1}}{a_{w-1}};\dots;\f{b_1}{a_1}\right], \quad
  \dots, \quad
  h_1 = \mathcal{H}\left[\f{b_1}{a1}\right], \quad
  h_0 = \mathcal{H}\left[\right] \equiv 1 \,.
\end{align}
Obviously, we have $(d/dx) h_{k} = f^{b_k}_{a_k} h_{k-1}$, $0<k\le w$, that is
the vector of functions ${\bf h} = (h_{w},\dots,h_0)^T$ obeys the following
differential equation:
\begin{align}
  \label{deCGPL}
  \frac{d}{dx}{\bf h} &= M(x)\,{\bf h} \,,
\end{align}
where the $(w+1)\times(w+1)$ matrix $M(x)$ has the form
\begin{align}  
  \label{matrixM}
  M(x) &=
  \left( {\begin{array}{cccccc}
      0 & f^{b_w}_{a_w} & 0 & \dots & 0 & 0\\
      0 & 0 & f^{b_{w-1}}_{a_{w-1}} & \dots & 0 & 0\\
      \dots & \dots & \dots & \dots & \dots & \dots \\
      0 & 0 & 0 & \dots & 0 & f^{b_1}_{a_1} \\
      0 & 0 & 0 & \dots & 0 & 0 \\
  \end{array}}
  \right) \,,
\end{align}  
and the rational functions $f^b_a(x)$ are given by Eq.~\eqref{eq:f-def}. 
The matrix in Eq.~(\ref{matrixM}) has simple poles at the zeros of the
polynomials $\Phi_n(x)$, which are the $n$-th roots of unity lying on the unit
circle $|x|=1$ in the complex-$x$ plane.
There exist exactly $w+1$ linearly independent solutions ${\bf v}_k$ to
Eq.~(\ref{deCGPL}).
It is useful to associate with this set of solutions the $(w+1)\times(w+1)$
matrix $W$ whose columns are formed by the vectors ${\bf v}_k$.
The determinant $\det W$ does not vanish at regular points of
Eq.~(\ref{matrixM}) indicating the linear independence of the solutions.
Moreover, the matrix $W$ is unique up to a constant matrix multiplier.

If, in the neighborhood of some point $x_0$, the matrix $M(x)$ has a
representation of the form
\begin{align}
  M(x) &= \frac{A}{x-x_0} + \sum_{k=1}^\infty (x-x_0)^{k-1} M_k\,,
  \label{eq:m}
\end{align}  
the fundamental solution $W$ can be found as a generalized series expansion
at this point in the form (see e.g.\ Ref.~\cite{gantmakher2000theory})
\begin{align}
  \label{Wexpansion}
  W(x) &= \left( \sum_{k=0}^\infty x^k U_k \right) x^A\,,
\end{align}
where $U_k$ are constant matrices given by the recursion relation
\begin{align}
  \label{Urecurrsion}
  U_0 = I\,, \qquad  U_n A - A U_n = \sum_{k=1}^{n} M_k U_{n-k} \,.
\end{align}
In particular, the solution of Eq.~(\ref{deCGPL}) around $x_0=0$ is given by
the product
\begin{align}
  \label{Wc}
  {\bf h}(x) = W(x){\bf c}\,,
\end{align}
where ${\bf c}$ is the constant vector $(0,0,\dots,0,1)^T$, which is determined
by the boundary conditions. 

In order to analytically continue ${\bf h}(x)$ away from $x=0$, we follow the
idea of Refs.~\cite{Lee:2017qql,Lee:2018ojn}.
We can match the solutions in different regions at some particular point that
belongs to both regions.
We find that, to cover the unit circle completely, we can evaluate $W(x)$ at
$x_0=0$ and at six other points placed symmetrically on the unit circle,
$x_k=e^{i\pi k/6}$, $k=0,1,\dots,5$.
We fix the boundary conditions at $x_0=0$ as described above.
The matchings to the six other expansions can be taken at the points
$\tilde{x}_k=\frac{1}{2}e^{i\pi k/6}$, i.e.\ in the middle of the straight lines
connecting $x_0$ and $x_k$.

The above algorithm has been realized in the {\it Mathematica} package
\verb+cyclogpl.m+.
It allows one to evaluate, with multiple-precision arithmetic, the cyclotomic
polylogarithms in Eq.~(\ref{ff}) with $b_j\le12$ and arbitrary weights.
An alternative numerical implementation of cyclotomic harmonic polylogarithms
is described in Refs.~\cite{Ablinger:2018sat,Ablinger:2018fut}.

\bibliography{veffSS} 

\end{document}